\documentclass[12pt]{article}
\usepackage{epsfig, bm, amsfonts, amssymb}
\usepackage{multirow}
\usepackage{tabularx}
\usepackage{hhline}
\usepackage[normalem]{ulem}
\usepackage{cite}
\usepackage[none]{hyphenat}
\usepackage[textsize=tiny]{todonotes}
\usepackage{chngcntr}
\counterwithin{figure}{section}
\counterwithin{table}{section}

\RequirePackage{color}

 \definecolor{MyDarkGreen}{rgb}{0.02,0.60,0.06}

\title{{\bf A Networks}-{\bf Science Investigation into the Epic Poems of Ossian}}
\author{ 
 {\it Joseph Yose}$^1$,
 {\it Ralph Kenna}$^{1,2}$,
 {\it P{\'{a}draig MacCarron}}$^3$, \\
 {\it Thierry Platini}$^{1,2}$ 
  and
 {\it Justin Tonra}$^4$ \\~\\
\begin{tabular}{ll}
  $^1$     & {\small{Applied Mathematics Research Centre, Coventry University, CV1 5FB, England}}\\
  $^2$     & {\small{${\mathbb L}^4$ Collaboration \& Doctoral College for the Statistical Physics of Complex Systems,}} \\       
	         & {\small{Leipzig-Lorraine-Lviv-Coventry, D-04009 Leipzig, Germany}} \\ 
	$^3$		 & {\small{Social \& Evolutionary Neuroscience Research Group, Department of Experimental}}\\
	         & {\small{Psychology, University of Oxford, OX1 3UD, England}}\\
  $^4$     & {\small{Discipline of English, School of Humanities, National University of Ireland,}} \\
	         & {\small{Galway, Ireland}}
\end{tabular}
{}\\~\\
}

\begin{document}
\maketitle
\vspace{-1cm}
{\Large
  \begin{abstract}
%
\noindent
In 1760 James Macpherson published the first volume of a series of epic poems which he claimed to have translated into English from ancient {Scottish-}Gaelic sources. 
The poems, which purported to have been composed by a third-century bard named Ossian, quickly achieved wide international acclaim. 
They invited comparisons with major works of the epic tradition, including Homer's \emph{Iliad} and \emph{Odyssey}, and effected a profound influence on the emergent Romantic period in literature and the arts. 
However, the work also provoked one of the most famous literary controversies of all time, colouring the reception of the poetry to this day. 
The authenticity of the poems was questioned by some scholars, while others protested that they misappropriated material from Irish mythological sources. 
Recent years have seen a growing critical interest in \emph{Ossian}, initiated by revisionist and counter-revisionist scholarship and by the two-hundred-and-fiftieth anniversary of the first collected edition of the poems in 1765. 
Here we investigate \emph{Ossian} from a networks-science point of view. 
We compare the connectivity structures underlying the societies described in the Ossianic narratives with those of ancient Greek and Irish sources. 
Despite attempts, from the outset, to position \emph{Ossian} alongside the Homeric  epics and to distance it from Irish sources, {our} results indicate significant network-structural differences between Macpherson's text and those of {Homer}.
They also show a strong similarity between Ossianic networks and those of the narratives known as {\emph{Acallam na Sen\'{o}rach}} (\emph{Colloquy of the Ancients}) from the Fenian Cycle  of Irish mythology.
                        \end{abstract} }
%
  \thispagestyle{empty}
%
%
  \newpage
%
                  \pagenumbering{arabic}

\section{Introduction}
\label{Introduction}

More than two hundred and fifty years ago, James Macpherson published what he claimed were translations into English of ancient Scottish Gaelic poetry dating from the ``most remote antiquity''~\cite{Macpherson1761,Macpherson1763,Macpherson1765,Macpherson1773}.
Now recognised as one of  ``the most important and influential works ever to have emerged''~\cite{Gaskill1996} from {Britain or Ireland,} \emph{Ossian} was also a profoundly significant work in its time, influential in the subsequent development of romantic nationalism in Europe. 
However, questions about its authenticity provoked one of the best-known controversies in literary history,  consequences of which are still to be found in  Ossianic studies today~\cite{Johnson1775,OHalloran1763,Lyons2007,Warner1762,Webb1763,OBrien1764,MacKenzie1805,Mulholland2009}. 
Revisionist scholarship since the 1990s brought the controversial episode  under academic scrutiny once again and  the topic has undergone something of a renaissance in the intervening years~\cite{Baer2012,Burnett2011,Curley2009,Meek1991,Fielding1996,Gaskill1991,StaffordinGaskill1996,Gaskill2004,Gaskill1996,Gaskill2008,Kidd1993,McNeil2007,Moore2000,Moore2004,Mulholland2009,Stafford1991,Womack1989,awforum}. 
Here we present a  fresh investigation into Macpherson's famous works, based upon a new quantitative approach~\cite{ourEPL}.
We contrast the social-network structures of \emph{Ossian} with particular works which have been identified as playing a potentially influential role in its composition: Homer's epics and Irish mythological texts. 
Our  aim is to determine whether or not the society described in Macpherson's work bears any structural similarity  to that of either corpus, from a networks perspective. 
We show that it is dissimilar to those of Homer but that there is a remarkable structural similarity  to the society underlying \emph{Acallam na Sen\'{o}rach} (\emph{Colloquy of the Ancients}) from the Fenian Cycle in Irish mythology.
This suggests a structural affinity between Macpherson's works and those whose authority he explicitly rejects in the commentary which accompanies his texts~\cite{Macpherson1761,Macpherson1763,Macpherson1765,Macpherson1773}.

In the next section of the paper we contextualise the origins of the Ossianic controversy, outlining the cultural circumstances in which the work appeared and accounting for the terms under which it received both acclaim and reproach. 
We discuss the work's significant cultural influence and  legacy, and examine its continued appeal as a fertile area of academic interest in the current revisionist and counter-revisionist contexts. 
Our main aim is addressed in Section~\ref{NetworkAnalysis} where we recall elements of network science and examine how network structures of \emph{Ossian} relate to Homeric and Irish literature. 
We summarise our conclusions in Section~\ref{Conclusions}.

\section{Contexts for {\emph{Ossian}}}
\label{Contexts}

In the mid-eighteenth century, the Highlands of Scotland were still culturally distinct from the rest of Britain and continental Europe, with a discrete societal structure, language, and customs~\cite{Womack1989}. 
It was the scene, in 1746, of the Battle of Culloden: the final confrontation of the Jacobite Rising which saw the defeat of Charles Edward Stuart by British forces loyal to the Hanoverian monarchy. 
The House of Stuart had ruled Scotland since the fourteenth century and the kingdoms of England and Ireland since the seventeenth, and their defeat was followed by the assimilation of the Highlands into  Great Britain and a vigorous campaign to undermine native Gaelic culture and to suppress Jacobite clans. 
Against this background of military defeat and cultural attenuation, the first fragments of a Scottish epic appeared. 
Macpherson's volume, {\emph{Fragments of Ancient Poetry, Collected in the Highlands of Scotland, and Translated from the Galic or Erse Language}}~\cite{Macpherson1760} was published in Edinburgh in 1760 and claimed to connect its readers to a vanished heroic age~\cite{StaffordinGaskill1996}. 
Framing the fragments is the blind bard Ossian, who relates narratives of battle, strife, and unhappy love from his aged solitude in the third century AD. 
The rhythmic prose and sparse diction of the poems created an evocative narrative style and ethereal atmosphere which captured the interest of a receptive public. 
Thus was launched the romantic portrayal of the Scottish Highlands which persists, in many forms, to the present day~\cite{Womack1989,McNeil2007}. 
The short volume promised to expand upon its fragmentary nature, with a ``hope that one work of considerable length, and which deserves to be styled an heroic poem, might be recovered and translated''~\cite{Macpherson1760}.

In 1760 and 1761, Macpherson made expeditions to the Scottish Highlands and Islands in search of traces of ancient Gaelic epic poetry, and in December 1761 (but dated 1762) the promised opus was published in London as {\emph{Fingal: An Ancient Epic Poem in Six Books, Together with several Other Poems composed by Ossian the Son of Fingal}}~\cite{Macpherson1761}. 
In March 1763, there followed {\emph{Temora: An Ancient Epic Poem in Eight Books, Together with several Other Poems composed by Ossian the Son of Fingal}}~\cite{Macpherson1763}. 
In 1765, the extant works were collected in the two-volume {\emph{The Works of Ossian, the Son of Fingal}} and accompanied by {\emph{A Critical Dissertation on the Poems of Ossian, the Son of Fingal}} by Hugh Blair, an influential Professor of Rhetoric at the University of Edinburgh~\cite{Macpherson1765}. 
A new edition, proclaiming to be ``[c]arefully corrected, and greatly improved''~\cite{Macpherson1773} was published in 1773 as \emph{The Poems of Ossian}. 
We base our analysis on the text of Howard Gaskill's modern scholarly edition, {\emph{The Poems of Ossian and Related Works}}, which is itself based on the 1765 edition~\cite{Gaskill1996}.
Certainly, a large number of variants between the 1765 and 1773 editions exist, but these consist primarily of a rearrangement in the order of the poems along with certain stylistic revisions.
Neither of these affects character interactions in the text, which are the  sources of data for our analysis. 
Throughout this paper we refer to the literary work as, simply, {\emph{Ossian}} (in italic font), with references to the character appearing as Ossian (in roman font).

 Macpherson's work has been interpreted, at least in part, as motivated by a desire to repair some of the injury inflicted on the Scottish Highlands in the aftermath of the Jacobite Rising~\cite{StaffordinGaskill1996,Gaskill2004}. 
The discovery of a work which testified to an ancient and noble Scottish literary heritage was a boon to scholars such as Hugh Blair, 
Adam Ferguson, and David Hume, and Blair's {\emph{Critical Dissertation}}, first published as a separate volume but incorporated into most editions of {\emph{Ossian}} after 1765~\cite{Macpherson1765}, played an important part in bolstering claims of its authenticity and augmenting its credibility. 
Active scholarly support for the Ossianic poems was motivated, to a certain degree, by a desire to forge an increased sense of national identity and unity in the face of the cultural fragmentation wrought by the Highland Clearances~\cite{Quint1993}.
Certainly, {\emph{Ossian}} was pivotal in popularising Highland literary traditions amongst audiences beyond the north of Scotland, prompting a vogue for its distinct formulation of cultural nationalism~\cite{McNeil2007}.

The wider impact of {\emph{Ossian}}  was enormous, its influence acknowledged by literary figures such as Blake, Byron, Coleridge, Goethe, Scott, and Wordsworth. Compositions by Brahms, Mendelssohn, and Schubert were also influenced by the work. 
Politicians and public figures reacted enthusiastically: Napoleon kept a copy in the portable library that accompanied him on his military campaigns, while Thomas Jefferson stated that Ossian  was ``the greatest poet that has ever existed''~\cite{Golden2002,Mulholland2009,Johnson2016}.  
Ossianic themes and figures featured in the works of painters such as Abildgaard, 
G{\'{e}}rard, 
Girodet,
Ingres,
Kauffmann, 
Krafft 
and Runge. 
Incorporated in 1820, but coming to worldwide prominence during the Civil Rights Movements of the 
1960s, Selma, Alabama owes its name to {\emph{The Songs of Selma}} in {\emph{Fingal.}}

{\emph{Ossian}} is widely acknowledged as having been instrumental in instigating a renewed interest in  national folklore, mythology, and poetry in the Europe of its time~\cite{Baer2012}. 
Interest in national literatures grew across the continent and {\emph{Ossian}} figures prominently in accounts of the growth of romantic nationalism during this period. 
Since it appeared to ``vindicate a fallen nation''~\cite{Quint1993} it had a particular resonance for countries subjected to conquest who looked to their ancient epics as reminders of a glorious national past~\cite{Baer2012,Quint1993}. 
Its romantic imagery was extended and adapted for literary nationalism emerging in other parts of Europe and it provided a model for articulating a non-classical iconography~\cite{Kidd1993}. 
{\emph{Ossian}} brought particular attention to Celtic works, provoking other nations  to undertake similar literary antiquarianism. 
Evans's {\emph{Some Specimens of the Poetry of the Ancient Welsh Bards}}~\cite{Evans1764} proved, in 1764, that Wales could boast a corpus of native poetry as interesting as that of the Highlands~\cite{Skene}. 
Thomas Percy's {\emph{Reliques of Ancient English Poetry}}~\cite{Percy1775} was published the following year, while Charlotte Brooke's {\emph{Reliques of Irish Poetry}}~\cite{Brooke1816} followed in 1789. 
The reception and development of the Ossianic tradition within various European contexts is explored in detail by B\"{a}r and Gaskill~\cite{Baer2012,Gaskill2004}.

For his supporters, comparisons between Macpherson's {\emph{Ossian}} and Homer's epics had already been established by the 1770s and the poems were seen to share an affinity with the {\emph{Iliad}} in particular~\cite{Gaskill2008,Michell2012}. 
Mitchell suggests that Macpherson was possibly assisted by Hugh Blair in providing comparative passages from Homeric epics in the footnotes to {\emph{Fingal}}. 
Such extracts have an authenticating function: they highlight similarities between episodes in Homer and {\emph{Ossian}}, and echo Blair's devotion of a large part of his {\emph{Critical Dissertation}} to establishing {\emph{Ossian}}'s epic character~\cite{Macpherson1765}. 
The comparison of Ossian to Homer is a central aspect of formalist analysis of the poetry and the phrase ``Homer of the North,'' which emerged in this context, is attributed to Madame de Sta{\"{e}}l in Ref.\cite{Michell2012} and is adopted by Gaskill~\cite{Gaskill2008}.
See Refs.\cite{McNeil2007,Gaskill2008,Mulholland2009,Baer2012} for further developments of this association.

Macpherson's work was not without its critics, however. 
These included Samuel Johnson -- the English author and scholar whose expertise, interests, and experience were embedded in the classical and neo-classical world.
The emphasis on native literature in Macpherson's poems provided a stark contrast which had greater resonance for the Romantic mode of literature than for Johnson's Augustan consciousness. 
He famously identified a ``strong temptation to deceit'' in the Ossianic poems~\cite{Johnson1775}.  
In 1773, Johnson journeyed into the western islands of Scotland, diligently inquiring about the sources of the Ossianic texts. 
His observations culminated in an argument that no ancient Scottish vernacular manuscripts were extant and a disbelief in the possibility of texts of {\emph{Ossian}}'s length passing intact through generations of oral transmission.  
However, Johnson's  reference to Gaelic as ``the rude speech of barbarous people''~\cite{Johnson1775}  
was indicative of the widespread English hostility towards Scotland and its culture in the post-Jacobite~age~\cite{Quint1993}.

Indeed, during the Imperial Era, British scholars and administrators aligned their attitudes towards classical history with imperialist ideologies. 
They came to view themselves as inheritors of the torch of civilisation, which had been passed from ancient Greece and Rome.
Classicists saw Homer as the epitome of this culture and his use of the epic form as a reflection of splendour of Greco-Roman civilisation.
Against this backdrop some viewed Scottish culture as inferior and uncivilised.
The conquest and colonisation of that country was therefore justified as representing a replacement of a  devalued, inferior and even expendable culture. 
The apparent discovery of an ancient epic tradition in the Highlands was a major challenge to these attitudes.

Thus, the dispute was  one of rivalry between emerging national identities within  Great Britain as well as an aesthetic clash between Classicism and Romanticism~\cite{Curley2009}.
Regardless of the accuracy of some of Johnson's claims about {\emph{Ossian}}, Macpherson's work would play a crucial formative role in the following decades as Classicism faded away in the face of an ascendant Romanticism.

Reaction to {\emph{Ossian}} in Ireland was no less vigorous, but proceeded from a different kind of objection.
Antiquarians suggested that the poems were corruptions of the tales of the Fenian Cycle  of Irish mythology.
Thinly veiled characters, places, and situations from the Irish epic tradition were identified, and further accusations of forgery were levelled at Macpherson. 
Sylvester O'Halloran characterised the nature of the response in 1763 by denouncing Macpherson's apparent attempt to misappropriate Ireland's Gaelic heroes for Scotland~\cite{OHalloran1763} 
(see also Refs.\cite{Lyons2007,Warner1762,Webb1763,OBrien1764}).
The Irish cause was further championed by the antiquarian and Gaelic scholar Charles O'Connor~\cite{OConnor1766}.
The character Ossian -- ``an illiterate Bard of an illiterate Age'' for Macpherson~\cite{OConnor1766} -- was  identified as Ois{\'{\i}}n, the warrior-poet of the Fenian Cycle in Irish mythology. 
Ossian's father, Fingal -- a third-century Scottish king, according to Macpherson -- was purportedly a version of Fionn mac Cumhaill, leader of a heroic band, the  Fianna {\'{E}}ireann.  
Cuchullin -- ``the General or Chief of the Irish tribes''~\cite{Macpherson1760} in Macpherson -- shares the name of  
C{\'{u}}~Chulainn, the hero of an entirely different epoch in Irish mythology.
The aggravation was worsened by Macpherson's inversion of the ancient relationship  between  Ireland and Scotland, including his reversal of the direction of migration of populations and folklore~\cite{StaffordinGaskill1996,Lyons2007}.  
O'Connor, claimed that Macpherson lacked ``decency, in the illiberal abuse of all ancient and modern writers, who endeavoured to throw lights upon the ancient state of Ireland, and North-Britain''~\cite{OConnor1766}.
Sylvester O'Halloran directly denounced {\emph{Ossian}} as an ``imposture'' and ``Caledonian plagery''~\cite{OHalloran1763}. 
Clare O'Halloran's work contains a more extensive analysis of Irish reactions to \emph{Ossian}~\cite{Halloran1989}.

The hotly contested debate about the origins and legitimacy of {\emph{Ossian}} had a discernible and enduring effect on ideas of Romanticism, sincerity and authenticity~\cite{Moore2004}.
While current literary scholarship on {\emph{Ossian}} has largely moved beyond the original terms of this debate,  aftershocks of this major literary-historical moment still reverberate~\cite{Curley2009}.

So is Ossian ``Homer of the North'' or an ``imposture''? 
Stated so bluntly the question is unanswerable, and reflects the entrenched polar positions of the early debate on the Ossianic controversy. 
While the majority of scholarship has  moved towards a more nuanced account of \emph{Ossian} than this question solicits, our understanding of the topics which inspire the question -- originality, authenticity, inspiration, allusion  -- can be enhanced by examining the structural foundations of \emph{Ossian} and its intertexts.
Here, we wish to add a new voice to the reassessment of the complex ideas at the heart of the Ossianic controversy by using the modern concept of social-network analysis. 
In particular, we take up the theme of the literary tradition from which it is claimed that {\emph{Ossian}}  emerged. 
As stated, two prime comparisons were made from the outset: to the Homeric epics on the one hand and to Irish mythology on the other, especially to the Fenian Cycle. 
Some previous studies focused on the identities and roles of individual characters in the narratives. 
Here we take a complementary approach; we identify the characters as nodes in a social network
and, instead of focusing on these individually, we address the totalities of relationships between them.
This is our motivation for comparing
 the social-network structure of  {\emph{Ossian}} to those of Greek and Irish narratives in Section~\ref{NetworkAnalysis}. 
However, before we do so, we first provide further context for the comparative texts.

The Homeric epics, the {\emph{Iliad}} and the {\emph{Odyssey}}, are considered to mark the beginning of the Western canon.
The {\emph{Iliad}} is dated to the eighth century BC, during the final year of the Trojan War. 
Its narrative recounts a quarrel between Agamemnon, king of Mycenae and leader of the Greeks, and Achilles, their greatest hero. 
The {\emph{Odyssey}}, in part a  sequel to the {\emph{Iliad}},  describes the journey home of the Greek hero Odysseus to his wife Penelope after the fall of Troy.

The Fenian Cycle  of Irish mythology mostly plays out in Leinster and Munster, the eastern and southern Irish provinces. 
The format is mainly poetic, with tales focusing on the adventures of  Fionn mac Cumhaill and the Fianna and there are also strong links with the Irish of Scotland. 
The most important source for the Fenian Cycle  is {\emph{Acallam na Sen{\'{o}}rach}} ({\emph{Colloquy of the Ancients}} or {\emph{Tales of the Elders}}). 
According to the stories, Ca{\'{\i}}lte  mac R{\'{o}}n{\'{a}}in and Ois{\'{\i}}n are the last  of the Fianna, and having survived into Christian times, they recount tales  of Fionn and his warriors to the recently arrived Saint Patrick. This, we are told, is how they came to be preserved in written form~\cite{DooleyRow1999}.

For our network analyses we use  Gaskill's version of {\emph{Ossian}} -- the first rendering of the  Ossianic corpus into a single volume~\cite{Gaskill1996}.
For the Homeric texts, we use Refs.\cite{Iliad} and~\cite{Odyssey} for the {\emph{Iliad}} and {\emph{Odyssey}}, respectively.
A relatively recent translation of {\emph{Acallam na Sen{\'{o}}rach}} is used~\cite{DooleyRow1999}.
To provide further contextualisation for our analysis, we also examine a more recent rendition of  parts of the Fenian Cycle  by Lady Gregory. 
(Specifically, we use {\emph{The Fianna}}, Part II of {\emph{Lady Gregory's Complete Irish Mythology}}~\cite{LadyGregory}.)
It is conceivable that other versions or translations of any of these texts may deliver differences  in respect of minor characters and relationships.
However, given the large numbers of characters and links between them in each case, the network properties are expected to be robust with respect to small variations in minor characters.

\section{Network Analysis}
\label{NetworkAnalysis}

Over the past couple of decades, network science has emerged as a powerful empirical tool to investigate complex systems. 
Real-world networks, such as those which appear in sociological systems, have certain topological features that render them different to  simple random graphs. 
For example, they tend to be {highly clustered} small worlds {with} community structures {whose} degree distributions are heavy tailed~\cite{Newman2003}. 
Here we briefly recall concepts {such as these} to mind before performing a network analysis of the societies depicted in {\emph{Ossian}} and comparative works.

Our approach to constructing the networks follows the methodology of Refs.\cite{ourEPL,ourEPJB}; 
characters, as well as the relationships and interactions between them, are identified by carefully reading  the texts with multiple passes through all of the material.
Characters are then represented by nodes.
All relationships and interactions identified are represented by  edges (links between the nodes).
Together the set of nodes and edges form a {\emph{network}}. 
Analysis of such networks delivers outputs in the form of statistics which can be considered as metrics.
We can also study the distributions of edges amongst the nodes of the various networks
as well as their connectivity information.
By comparing these we can gauge similarities and dissimilarities between the social structures underlying the narratives.

We begin by comparing some standard topological measures of the networks underlying each text. 
We then analyse their degree distributions and finally we look at the spectral distance from the adjacency matrices of the graphs.

\subsection{Network topology}
\label{NT}

We consider a network with $N$ {\emph{nodes}} and $M$ {\emph{edges}}, each of which links a pair of nodes.  
In principle, when the feature which connects nodes is oriented, one may consider the edges as being {\emph{directed}} but for the level of detail we consider here, undirected networks are appropriate.
Indeed, following earlier network analyses of narrative texts, we  distinguish between edges which represent {\emph{positive (or friendly) and {\emph{negative}} (or hostile) relationships between characters~\cite{ourEPL,ourEPJB}. 
Positive links are established between pairs of  characters that are related; communicate directly with one another; speak about one another or are present together where it is clear that they know each other. 
Negative links are formed when  characters meet in  conflict (a positive link is not made if they speak only here); or when animosity is explicitly declared by one character against another and it is clear they know each other.}}
This level of detail renders the relationships we wish to consider reciprocal and maximises sample sizes and hence statistical power.
The {\emph{degree}} $k_i$ of node $i$ is {then defined as} the number of edges {connected to} that node and its maximum value over the entire network is denoted $k_{\rm{max}}$.
The mean degree of the network is denoted by $\langle{k}\rangle$ and is given by $2M/N$.

A {\emph{geodesic}} is path of  minimum length between two nodes of the network  along a sequence of edges.
The characteristic {\emph{path length} $\ell$ is the mean geodesic distance on the network.

{The number of potential edges between the $k_i$ neighbours of node $i$ is $k_i(k_i-1)/2$. 
If the actual number of links between that set of neighbours is $n_i$, then the {\emph{clustering coefficient}}  $C_i = 2n_i/k_i(k_i-1)$  
{is the density of interactions between neighbours of node~$i$.}
The mean of these quantities taken over all nodes is the {\emph{clustering coefficient}}
 of the network and denoted by $C$.
It quantifies the degree to which nodes in the network tend to cluster together. 
}

Another measure of clustering, sometimes called the transitivity,  is the proportion {{$C_T$}} of connected triplets which are closed. 
If there are edges between the pair of nodes $(i, j)$ and the pair $(j, k)$ then {$C_T$}  measures the probability of there also being an edge between $i$ and $k$.
It quantifies the extent to which a neighbour of a neighbour is a neighbour.

We frequently wish to compare networks to a random one of the same size $N$ and same average degree $\langle{k}\rangle$. 
For random networks of the Erd{\H{o}}s-R{\'{e}}nyi type, the path length and clustering are~\cite{Fronczak,albert2002statistical,NewmanPark,PMCthesis}
\begin{equation}
\ell_{\rm{rand}} \approx \frac{{\ln{N}-\gamma_{\rm{E}}}}{\ln{\langle{k}\rangle}} + \frac{1}{2}
\quad
{\mbox{and}} 
\quad
C_{\rm{rand}} \approx \frac{\langle{k}\rangle}{N}.
\end{equation}
Here $\gamma_{\rm{E}} \approx 0.5772$  is  the Euler-Mascheroni constant and
the expression for $C_{\rm{rand}}$ holds as an approximation for both the clustering coefficient $C$ and the transitivity $C_T$.
{If a network has} $\ell \approx \ell_{\rm{rand}}$ and $C,C_T \gg C_{\rm{rand}}$,  the network is {\emph{small world}}~\cite{WattsStrogatz}. 
This means that, although most nodes of the network are not nearest neighbours of each other, it takes only a small number of hops or steps along edges to connect them. 
Many social networks exhibit this property; strangers tend to be connected by a short chain of relationships.

\begin{figure*}[t]
\begin{center}
\includegraphics[width=1.0\columnwidth,angle=0]{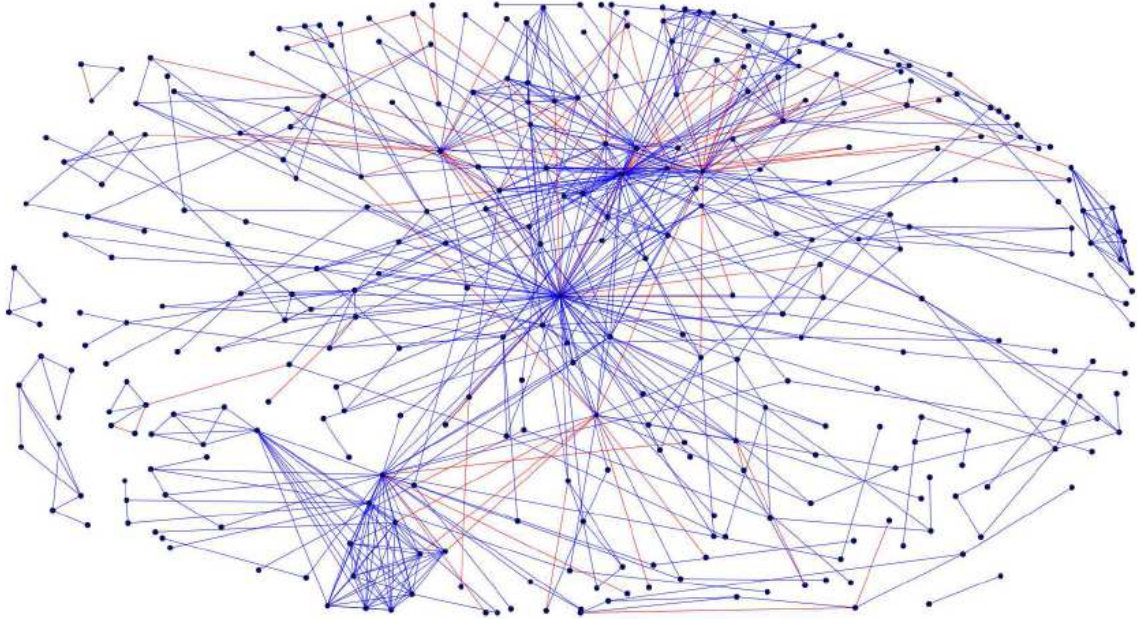}
\caption{The entire network of 748 relationships between 325 characters depicted in MacPherson's {\emph{Ossian}} (Gaskill's version). Positive (friendly) and negative (hostile) edges are indicated in blue and red, respectively. 
}
\label{Fig3p1}
\end{center}
\end{figure*}

Another property, commonly measured for social networks, is the {\emph{assortativity}} $r$~\cite{NewmanPark}. 
This is the tendency for similar nodes of a network to attach to each other~\cite{Newman2002}.
The degree assortativity, in particular, measures the extent to which nodes are likely to interact with other nodes of similar degree.
It is measured using the  Pearson  coefficient $r$.
If this is positive, the graph is said to be {\emph{assortative}} and if it is negative the graph is called {\emph{disassortative}}.

In the full network, there is a  tendency to disfavour odd numbers of negative links in a closed triad.
This  is related to the notion of {\emph{structural balance}}~\cite{Heider}.
One way to quantify this is by introducing  the statistic $\Delta$ as the percentage of closed triads that contain an odd number of    positive links.
Large values of $\Delta$ capture the notion that ``the enemy of my enemy is my friend''; hostility between two characters is suppressed if they have a common foe.
Clearly $\Delta$ is only relevant to the complete network rather than its positive or negative subsets because for the former $\Delta = 1$ by definition, while in the latter it is zero.

A {\emph{connected component}} of a network is a sub-graph, all nodes of which are connected to each other by sequence of edges, and which is not connected to vertices in the remainder of the network.
In theory, the smallest possible component is an isolated node and the biggest is the entire network itself, provided that every node can reach every other along a chain of edges. 
In most practical situations, however, a network has many components and these have a range of sizes. 
One is then primarily interested in the {\emph{giant component}}, namely the one with the greatest number of nodes.
The giant component is in itself a connected network. 
We represent by $G_c$ the proportion of network nodes which are part of the giant component.

\begin{table}[!b]
\caption{Properties of the full, positive and negative networks for {\emph{Ossian}} and comparative texts. 
Here, ``{\emph{Acallam}}'' means {\emph{Acallam na Sen{\'{o}}rach}} and ``{\emph{Gregory}}'' stands for  Lady Gregory's text. 
}
\begin{center}
\resizebox{\textwidth}{!}{%
\begin{tabular}{|l|l|c|c|c|c|c|c|c|c|c|c|c|c}  \hline
        & Narrative         & $N$ &$M$ 
                                       &${\langle{k}\rangle}$
																							& {$\ell$} 
																							       & {$ \ell_{\rm{rand}}$}
																									          & {$C$} & {$C_{\rm{rand}}$} & {$C_T$} &$r$    & $\Delta$ &$G_c$~~           \\  \hline  
        & {\emph{Ossian}}   & 325 & 748& 4.60 & 3.62 & 3.91 & {0.49}& {0.01} & 0.27& -0.08 & 0.95    & $88.62 \%$     \\ 
\multirow{4}{*}{\rotatebox{90}{Full}}
        & {\emph{Acallam}}  & 732 &1606& 4.39 & 3.79 & 4.57 & {0.37}& {0.01} &0.19 & -0.10 & 0.97    & $76.91 \%$     \\
        & {\emph{Gregory}}  & 355 & 913& 5.14 & 3.10 & 3.73 & {0.44}& {0.01} &0.16 & -0.18 & 0.97    & $94.65 \%$     \\
        & {\emph{Iliad}}    & 694 &2684& 7.74 & 3.49 & 3.42 & {0.44}& {0.01} &0.45 & -0.08 & 0.98    & $99.42 \%$     \\
        & {\emph{Odyssey}}  & 301 &1019& 6.77 & 3.29 & 3.18 & {0.45}& {0.02} &0.38 & -0.08 & 0.97    & $98.34 \%$     \\
\hline
        & {\emph{Ossian}}   & 309	& 666& 4.31 & 3.65 & 4.03 & {0.42}& {0.01} &0.31 & -0.06 &         & $82.20 \%$     \\ 
\multirow{4}{*}{\rotatebox{90}{Positive}}
        & {\emph{Acallam}}  & 722 &1513& 4.19 & 3.72 & 4.69 & {0.38}& {0.01} &0.20 & -0.09 &         & $71.33 \%$     \\
        & {\emph{Gregory}}  & 337 & 833& 4.94 & 3.23 & 3.78 & {0.45}& {0.01} &0.18 & -0.17 &         & $91.99 \%$     \\
        & {\emph{Iliad}}    & 640 &2329& 7.28 & 3.80 & 3.47 & {0.44}& {0.01} &0.58 & 0.02  &         & $85.94 \%$     \\
        & {\emph{Odyssey}}  & 299 & 989& 6.62 & 3.42 & 3.21 & {0.45}& {0.02} &0.40 & -0.08 &         & $97.32 \%$     \\
\hline
        & {\emph{Ossian}}   & 87 &  82 & 1.89 & 5.30 & 6.61 & {0.00}& {0.02} &{0.00}& -0.31 &         & $ 70.11\%$     \\ 
\multirow{4}{*}{\rotatebox{90}{Negative}}
        & {\emph{Acallam}}  &  86 &  93& 2.16 & 2.32 & 5.53 & {0.00}& {0.00} &{0.00}& -0.30 &         & $24.42 \%$     \\
        & {\emph{Gregory}}  & 95  &  80& 1.68 & 4.75 & 8.13 & {0.00}& {0.01} &{0.00}& -0.30 &         & $45.26 \%$     \\
        & {\emph{Iliad}}    & 321 & 355& 2.21 & 4.46 & 7.00 & {0.00}& {0.00} &{0.00}& -0.45 &         & $ 90.34\%$     \\
        & {\emph{Odyssey}}  &  41 &  30& 1.46 & 1.88 & 8.74 & {0.00}& 0.04 & {0.00}& -0.18 &         & $26.83 \%$     \\  \hline
\end{tabular}
}
\end{center}
\label{table1}
\end{table}

These statistics, together with the degree distribution described below, encapsulate a variety of characteristics of networks.
By comparing them between different networks, we may obtain {quantitative indications of 
similarities and differences between them.} 
In previous studies, the literary usage of such an approach was comparison and classification of epic narratives and sagas~\cite{ourEPL,ourEPJB}.
Our objective here is to measure these statistics for the social  network underlying {\emph{Ossian}} and those from texts to which Macpherson's has been compared. 
Our aim is to use these to formulate quantitative comparisons. 
However, another use of networks is to aid visualisation and to this end, we depict the entire {\emph{Ossian}} network in Figure~\ref{Fig3p1}.
In Appendix~\ref{AppendixA} we list some of the most important characters of this network and of its positive and negative sub-graphs.

In    Table~\ref{table1}, we list some network statistics for {\emph{Ossian}} and comparative texts.
The statistics  indicate that, for each narrative, the full networks have quite similar properties to the positive ones and are quite different to their  negative counterparts.
We interpret this as  reflecting that, although conflict is an important element of each story, 
the networks properties are  dominated by positive social interactions.  
Indeed, although {\emph{Ossian}} is generally seen as a rather morose and melancholy work, with 666 positive edges between 309 nodes, and 82 negative edges between 87 nodes, the former dominate the  Ossianic system. 
Strong degrees of disassortativity signal that the negative networks are held together by hubs of highly connected characters in conflict with multitudes of lesser significant ones. 
Short average path lengths when compared to random networks {and} the absence of clustering  mean they are not  small-world.
Thus the negative networks are not reflective of genuine societies; 
the similarities between their relatively trivial topologies do not allow them to deliver information that may be used to categorise or differentiate between the different narratives.

{
Like many complex  networks, the full and positive societies listed in Table~\ref{table1} are  structurally balanced small worlds with non-trivial topologies  and large proportions of nodes belong to their giant components~\cite{Newman2003,albert2002statistical}. 
The statistics therefore capture universal properties of mythological networks (see Refs.\cite{ourEPL,ourEPJB}) and invite comparison. 
We observe, for example,  that the average degrees of \emph{Ossian} are more similar to those of  \emph{Acallam} and Lady Gregory's text than they are to those of the Homeric epics.
However, although the path lengths of {\emph{Ossian}} are close to those  of  \emph{Acallam}, 
they are also close to those of the {\emph{Iliad}} and rather different to those of Lady Gregory's text, 
which likewise has quite different values of assortativity. 
The various clustering coefficients also fail to distinguish similarities. 
Therefore, although the network statistics are informative in a broad sense, 
they  do not  signal whether the Ossianic networks are more similar to those of the Irish or the Classics. 
To  investigate further we turn to the  degree distributions.
}

\subsection{Comparison of Degree Distributions}
\label{CDD}

\begin{figure*}[t]
\begin{center}
\includegraphics[width=0.49\columnwidth, angle=0]{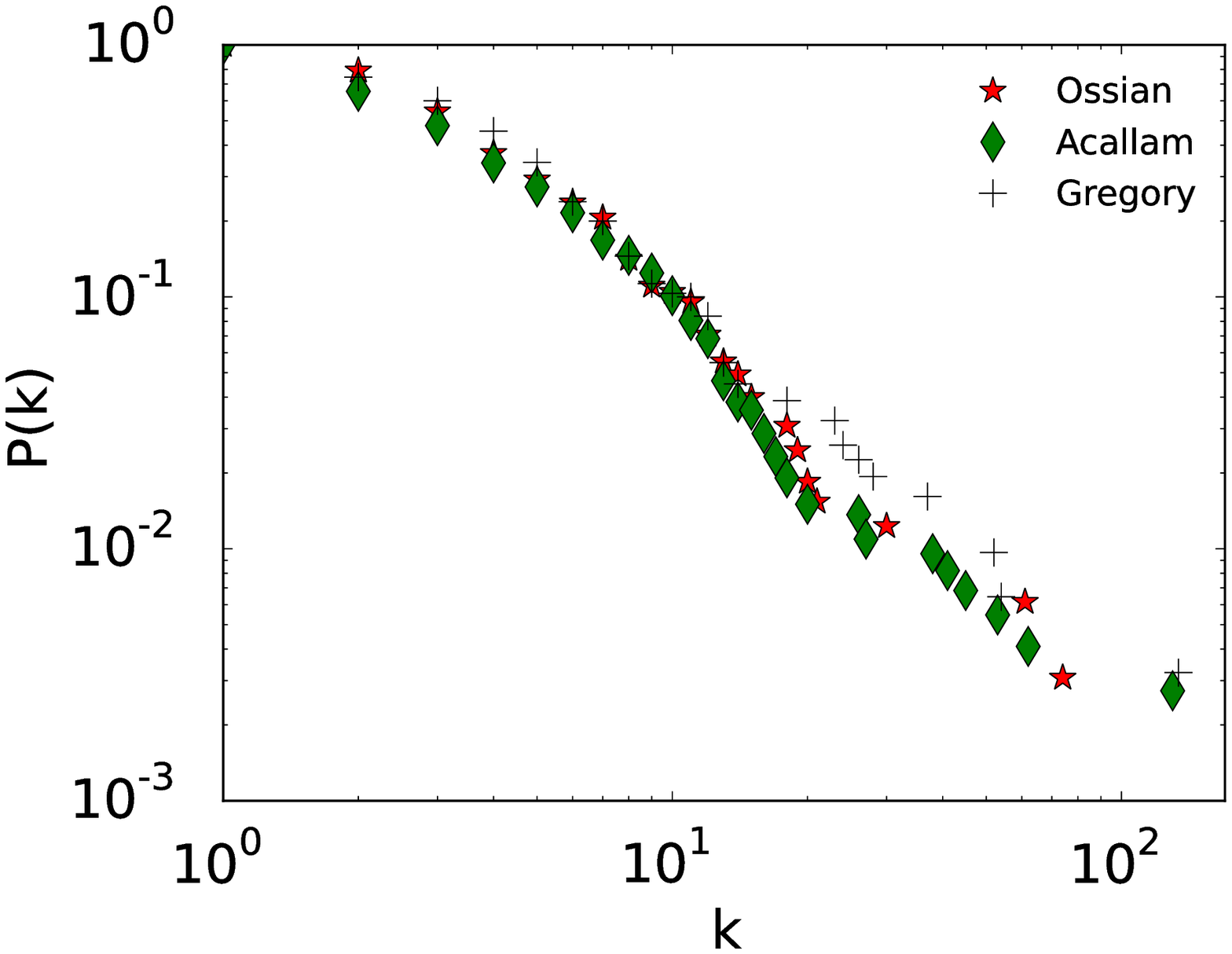}
\includegraphics[width=0.49\columnwidth, angle=0]{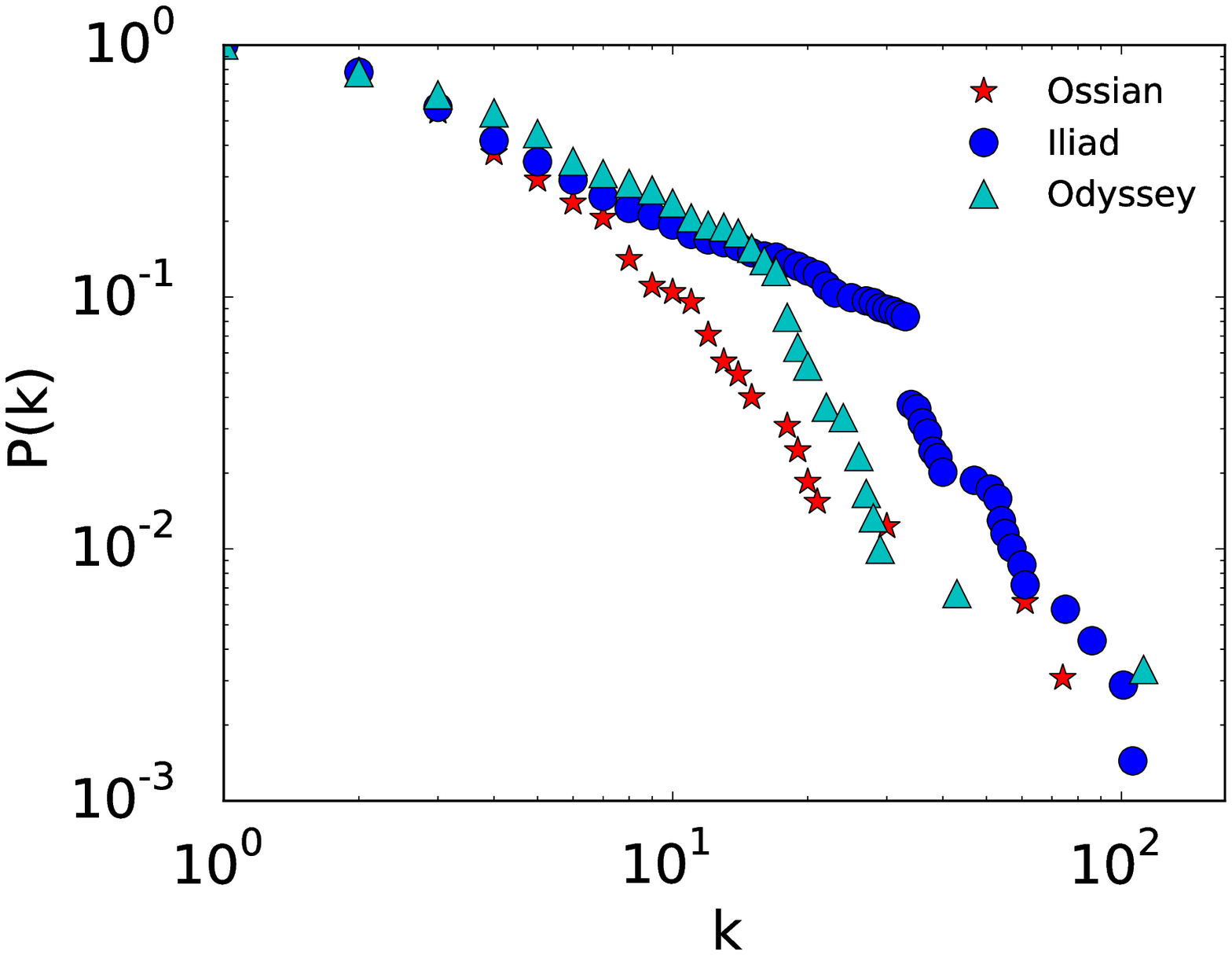}
\caption{The complementary cumulative degree distributions of the full networks indicate that the society depicted in {\emph{Ossian}} more closely resembles those of the Irish {\emph{Acallam na Sen{\'{o}}rach}} and  Lady Gregory's text (denoted here by ``{\emph{Gregory}}'') than the {\emph{Iliad}} or {\emph{Odyssey}}.
}
\label{Fig3p2}
\end{center}
\vspace{0.3cm}
\end{figure*}

The degree distribution $p(k)$ of a network governs the probabilities for nodes to have certain numbers of edges.
Complex networks tend to be right-skewed reflecting the small probability for a node to have a high degree and vice versa~\cite{Newman2003}. 

\subsubsection{Functional Forms}
\label{FFFF}

The functional form of the decay gives us further information about the structure of the connectivity of the network.
Rather than displaying the probability distribution functions themselves, the convention is to present the complementary cumulative degree distributions $P(k)$ because it reduces the noise in the tail~\cite{Newman_power};
$P(k)$ represents the probability that the degree  of a node is greater than or equal to the value~$k$. 

The cumulative distributions for the  networks are plotted in Figure~\ref{Fig3p2}. 
In the  panels, the degree distribution for the full {\emph{Ossian}} network 
(all components and including both positive and negative edges)
is compared to those of the Irish  (left) and  Greek (right) texts. 
(See Appendix~\ref{AppendixB} for equivalent plots restricted to the giant components of the positive networks.)
Clearly the distribution for {\emph{Ossian}} more closely resembles those from the Irish corpus.
To explore further, we fit different standard functions to these various distributions.
{In section \ref{KSC}, we}  also apply  the Kolmogorov-Smirnov test to investigate the similarities between degree distributions in a parametric-independent manner.

Small-world networks usually fall into one of a number of classes of degree distributions~\cite{Amaral}. 
The basic power-law degree distribution takes the form 
$ p(k) \sim k^{-\gamma}$.
This distribution lacks a scale and networks which are so described are referred  to  as  {\emph{scale free}}. 
Such power-law behaviour is ordinarily found only above some minimum degree \ $k_{\rm{min}}$~\cite{Clauset2009}.   
We need not consider $k_{\rm{min}}=0$  because nodes with this degree are disconnected from the network. 
The next lowest possible value $k_{\rm{min}}=1$  can lead to 
statistics uncharacteristic of networks~\cite{PMCthesis}. 
Since we wish to capture and characterise as many data as possible, when analysing the degree distributions we use~$k_{\rm{min}} = 2$.

Often networks have a cut-off beyond which the simple power law  no longer describes their degree distributions~\cite{albert2002statistical}. 
Such a cut-off sets a scale and, if it occurs at $k = \kappa$, it can be described by the truncated form
\begin{equation}
 p(k) \sim k^{1-\gamma}e^{-k/\kappa}.
 \label{tpl}
\end{equation}
We will find that the social network underlying the {\emph{Iliad}}   is best described by a this type of distribution~\cite{ourEPL}.

The log-normal distribution has the form
\begin{equation}
 p(k) \sim \frac{1}{k} \exp{\left[{-\frac{(\ln{k}-\mu)^2}{2\sigma^2}}\right]},
\label{log-normal}
\end{equation}
where $\mu$ and $\sigma$ represent the mean and standard deviation, respectively, of the natural logarithm of the degree variable. 
These also set a scale, so such distributions are not scale free.  
We will see that {\emph{Ossian}}, {\emph{Acallam na Sen{\'{o}}rach}} and Lady Gregory's text are best described by log-normal distributions.

Finally we consider distributions given by 
\begin{equation}
p(k) \sim 
\left({\frac{k}{\kappa}}\right)^{\alpha}
\exp{\left[{-\left({\frac{k}{\kappa}}\right)^\beta}\right]}.
\label{Wei}
\end{equation}
If $\alpha = 0$ this is a stretched exponential,  if $\alpha = \beta -1$ it is the Weibull distribution and if $\alpha = 0$ and $\beta =1$ this is just the standard exponential distribution.
The stretched exponential is actually the complementary cumulative function of the Weibull distribution.
Again the parameter $\kappa$ sets a scale in both cases.
We will find that the social network underlying the {\emph{Odyssey}} is best described by a Weibull distribution.

\begin{table}[!b]
\caption{Relative minimum-information-loss probabilities for the various degree distribution functions of the full networks underlying  the considered narratives. 
The probabilities for the most likely distribution for a given narrative are highlighted in boldface.
}
\begin{center}
\begin{tabular}{|l|c|c|c|c|c|} \hline 
                     & {Power law} & Truncated         & Log normal       & {Exponential} & Weibull     \\
                     &             & power law         &                  &               &             \\
\hline
{\emph{Ossian}}      & {$\sim 0$}  & {0.11}        &  {\bf{0.54}}    &   {$\sim 0$}  &    {0.35}        \\
{\emph{Acallam}}     & {$\sim 0$}  &   $\sim 0$        &  {\bf{0.97}}  &   {$\sim 0$}  &  0.03      \\
{\emph{Gregory}}     & {$\sim 0$}  &   $\sim 0$        &  {$\sim$ {\bf{1}}}        &   {$\sim 0$}  &  {$\sim 0$}       \\
{\emph{Iliad}}       & {$\sim 0$}  &   $\sim$ {\bf{1}}  & $\sim 0$         &   {$\sim 0$}  &    $\sim 0$  \\
 {\emph{Odyssey}}    & {$\sim 0$}  &   0.32            &     0.16         &   {$\sim 0$}  &    {\bf{ 0.52 }}     \\ 
 \hline 
\end{tabular}
\end{center}
\label{table2}
\end{table}

Here we fit data for each narrative to each of the above functional forms. 
We use maximum likelihood estimators to determine the parameters for each probability distribution as described in Ref.\cite{Clauset2009}.
To determine which of these models are most likely, we use the Akaike information criterion (AIC )~\cite{AIC}.
If $q$ parameters are estimated using a model, this information-theoretic quantity is defined as $ \mathrm{AIC} = 2q - 2\ln{\mathcal{L}}$ where  $\mathcal{L}$ represents the  likelihood for the parameters entering the probability distribution. 
Thus the higher the likelihood, the smaller  the AIC value. 
The AIC penalises usage of too many estimated parameters to compensate for the fact that increasing these tends to improve the goodness of fit.
One may introduce a correction factor to take into account the finiteness of sample sizes. The corrected AIC is 
\begin{equation}
 \mathrm{AIC}_{\rm{c}} = \frac{2qn}{n - q - 1} - 2\ln{\mathcal{L}} ,
\end{equation}
where $n$ is the sample size.
The corrected formula therefore penalises small values of $n$. 
(Although this formula assumes that the model is univariate, linear, and has normally-distributed residuals, it may often be used more generally in place of a more precise correction~\cite{Burnham}.)
To chose amongst the various models, 
one forms $\exp{[({\rm{AIC}}_{{\rm{c}};\rm{min}} - {\rm{AIC}}_{\rm{c}})/2]}$ where AIC$_{{\rm{c}};\rm{min}}$ is the minimum AIC$_{\rm{c}}$ value for a given narrative.
These can be interpreted as the relative probability that a given model minimises the estimated information loss coming from using that model instead of the ``true'' distribution function which governs the data. 
These are listed for each network in Table~\ref{table2}.
We also present the results of the maximum-likelihood fits to the various {\emph{Ans{\"{a}}tze}} in Table~\ref{table3}.

\begin{table}[!b]
\caption{Maximum-likelihood estimates for the various parameters associated with the probability distributions fitted to the data for the full networks. }
\begin{center}
\resizebox{\textwidth}{!}{%
\begin{tabular}{ |l                         |c          |c         |c                   |c          |c                   |c             |c       |c|}    \hline
                                            & Power     &\multicolumn{2}{c|}{Truncated} & \multicolumn{2}{c|}{Log normal}&Exponential &\multicolumn{2}{c|}{Weibull }  \\
                                            & law       &\multicolumn{2}{c|}{power law} & \multicolumn{2}{c|}{          }&   & \multicolumn{2}{c|}{          } \\\hline
                                            & $\gamma$  & $\gamma$ &$\kappa$            & $\mu$     & $\sigma$           & $\kappa$      & $\beta$ & $\kappa$ \\
\hline
{\emph{Ossian}}  & {2.1(7)} & 1.4(1)  &   {15.7(6)}  &  {{\bf{0.5(7)}}}  &   {{\bf{1.2(0)}}}  &  4.0(1)		& {0.5(2)} &  { 0.6(6)}  \\\hline
{\emph{Acallam}} & {2.0(5)} & 1.4(1)  &  {18.9(4)}  &  {\bf{0.8(5)}}  &   {\bf{1.1(8)}}  &  4.7(1)    & {0.5(1)}  &   {0.8(4)}     \\\hline
{\emph{Gregory}}  & {1.9(6)} & 1.2(1)  &   {17.3(6)}   &  {\bf{1.2(6)}}  &   {\bf{1.0(3)}}   &   5.2(1)   & {0.5(2)}  &   {1.3(4)}     \\\hline
{\emph{Iliad}}   & {1.8(4)} & {\bf{1.4(0)}}  &   {\bf{39.6(3)}}   &  {0.0(8)}  &  {1.8(0)}   &  8.1(1)    & 0.4(1)  &   {0.5(6)}     \\\hline
{\emph{Odyssey}} & {1.8(6)} & {0.7(1)}  &   {12.2(12)}   & {1.5(7)}  &   {1.0(2)}  &  {6.9(2)}   & {\bf{0.7(4)}}  &   {\bf{4.3(2)}}     \\
 \hline 
\end{tabular}
}
\end{center}
\label{table3}
\end{table}

The results indicate that, of the various candidate models for the degree distributions, 
the  networks underlying {\emph{Ossian}}, {\emph{Acallam na Sen{\'{o}}rach}},  and Lady Gregory's text are most probably log-normal
with parameters  {{$\mu$ between  {0.5} and 1.2}} and $\sigma$ {{about 1 or a little above}}. 
(However, in the case of {\emph{Ossian}}, a Weibull distribution cannot be ruled out.)
A truncated power law with  {{$\gamma \approx 1.4$ and $\kappa \approx 39.6$}} is  most probable for the {\emph{Iliad}} network and the Weibull distribution with {{ $\beta \approx 0.7$ and $\kappa \approx 4.3$}} is most probable for the {\emph{Odyssey}}.
(However, truncated power-law and log-normal distributions cannot be dismissed for the \emph{Odyssey}.) 
In Appendix~\ref{AppendixB} we give the relative probabilities and maximum-likelihood parameter estimates for the giant components of the positive networks.

\subsubsection{Kolmogorov-Smirnov Comparisons}
\label{KSC}

The AIC does not deliver information about the quality of a model in an absolute sense - only on relative quality.
As an alternative, we  next  investigate similarities between the degree distributions associated with different narratives in a non-parametric manner.
The  Kolmogorov-Smirnov test  does not  require knowledge of the {distributional forms} and,
although it was developed for continuous distributions,  can be used in the discrete case provided one is interested in comparing two samples to each other rather than comparing one sample to a probability distribution function~\cite{conover1972kolmogorov,conover1980practical, daniel1990applied}.
This is the situation that pertains in the present instance.
If $f(k) $ and $g(k)$ are the two empirical distribution functions, the test statistic is $D = {\rm{sup}}|f(k) - g(k)|$.
The null hypothesis is that the samples are drawn from the same underlying distribution. 
Assuming this, the $p$ value gives the probability that the two distributions are as different as observed.
If the $p$ value is large ($p>0.05$), we cannot reject the null hypothesis and if it is small ($p<0.05$), we may conclude that the two samples are likely to have come {from} different populations with different distributions.
In the current context, we interpret large $p$ values as indicating similarities between the degree distributions of the networks underlying two narratives, a circumstance we describe as a ``match''. Small $p$ values suggest dissimilarities. 
The results for the full and positive networks {{(considering all components in each case)}} are presented in Table~\ref{table4}. 
(We are not interested in similarities between hostile networks as these reflect their trivial topologies rather than interesting societal structure. Indeed we checked, and the negative networks from every text analysed matches every other, indicating another universal, non-discriminating feature of such narratives.)

\begin{table}[!b]  
\caption{Results, in terms of $p$ values, of the Kolmogorov-Smirnov tests applied to the full and positive networks (when they differ, data for the latter are given in square brackets).
The large values of $p$ (highlighted in boldface) in the cell corresponding to the {\emph{Ossian}} column and {\emph{Acallam}} row suggest similarities in the corresponding degree distributions. The small values in other cells suggest the absence of similarities between the 
corresponding character networks.
 }
  ~\\
  \centering
  \begin{tabular}{clllll}
\cline{1-2} 
\multicolumn{1}{|c|}{\textbf{}} & \multicolumn{1}{l|}{{\emph{Ossian}}}          &                                     &                                       &                                      &  
 \\ \cline{2-3}
\multicolumn{1}{|c|}{{Full}}           & \multicolumn{1}{l|}{{\bf{0.79} [{\bf{0.95}}]}}& \multicolumn{1}{l|}{{\emph{Acallam}}} &                                       &                                      & 
 \\ \cline{2-4}
\multicolumn{1}{|c|}{{[Positive]}}     & \multicolumn{1}{l|}{{$<10^{-2}$ [{0.04}]}} & \multicolumn{1}{l|}{{0.03 [0.02]}}  & \multicolumn{1}{l|}{{\emph{Gregory}}}  &                                      & 
 \\ \cline{2-5}
\multicolumn{1}{|c|}{{Networks}}       & \multicolumn{1}{l|}{{{$<10^{-5}$}}}           & \multicolumn{1}{l|}{{$<10^{-5}$}} & \multicolumn{1}{l|}{{$<10^{-2}$}} & \multicolumn{1}{l|}{{\emph{Iliad}}}  &  
 \\ \cline{2-6}
\multicolumn{1}{|c|}{\textbf{ }}              & \multicolumn{1}{l|}{{$<10^{-3}$}}           & \multicolumn{1}{l|}{{$<10^{-3}$}} & \multicolumn{1}{l|}{{$<10^{-3}$}} & \multicolumn{1}{l|}{$\le10^{-2}$} & \multicolumn{1}{l|}{{\emph{Odyssey}}} \\ 
\cline{2-6}
\hline   
\end{tabular}
\label{table4}
\end{table}

It is clear from the table that whether we use the full network or positive sub-network, there are strong matches between the social structures of  {\emph{Ossian}} and {\emph{Acallam na Sen{\'{o}}rach}}.
The test fails to detect  a match between {\emph{Ossian}} and either of the Homeric texts.
In Appendix~\ref{AppendixB} we apply  similar tests to the giant components of the various networks. 
These detect matches between {\emph{Ossian}} and both of the Irish texts as well as between {\emph{Acallam na Sen{\'{o}}rach}} and Lady Gregory's version.
Again, however, they detect no match between {\emph{Ossian}} and the Homeric networks. (See Table~\ref{tableB4} of Appendix~\ref{AppendixB}.)

~\\
{\subsection{Spectral Distances}}
\label{spectrallll}

As well as comparing structural properties of {{the societies underlying the various narratives through network statistics and degree distributions, one may}} compare two networks directly.
The adjacency matrix of a network contains all of its connectivity information and one way to compare these  is 
 to use the Ipsen-Mikhailov (IM) distance~\cite{Ipsen}. 
Although  originally defined for  dynamical biological networks, it is quite robust and can be considered wherever a quantitative comparison between networks is needed~\cite{Jurman}.

\begin{table}[!b]
\caption{ Spectral distances from the networks in {\emph{Ossian}} to those of the other texts. Lower values indicate a greater degree of similarity between the graphs.}
\begin{center}
\begin{tabular}{|l|c|c|c|c|} \hline 
Distance from {\emph{Ossian}}  & {\emph{Acallam}} & {\emph{Gregory}}  & {\emph{Iliad}} & {\emph{Odyssey}} \\ \hline 
 Full network                  & 0.03             & 0.02             & 0.07           & 0.09 \\ \hline
 Positive sub-graph             & 0.03             & 0.03             & 0.08           & 0.10  \\ 
\hline 
\end{tabular}
\end{center}
\label{table5}
\end{table}

The adjacency matrix {$A$} of a network is constructed so that element $A_{ij}$ is one if there is an edge from node $i$ to node $j$ and zero otherwise.
All diagonal elements are zero because there are no edges from a node to itself (loops) in the simple graphs considered here.
The matrix $A$ is symmetric since our networks are undirected.
We  construct the network's so-called Laplacian matrix $L$ by defining its elements as $L_{ij}=\delta_{ij}k_i-A_{ij}$ where $\delta_{ij}$ is the Kronecker delta which is 1 if $i = j$ and 0 otherwise. Each diagonal element of the Laplacian gives the degree of the associated node.
This Laplacian matrix has $N-1$ eigenvalues $\lambda_i$~\cite{Newman_book}.
The spectral function  $\rho(\omega)$ is a sum of Cauchy-Lorentz distributions about the square-root of the eigenvalues that is defined as
\begin{equation}
\rho(\omega) = K \sum_{i=1}^{N-1} \frac{s}{(\omega-\sqrt{\lambda_i})^2 + {s}^2},
\end{equation}
where $K$ is a normalisation constant defined by $\displaystyle{\int_{0}^{\infty}{ \rho(\omega) d\omega} =1}$. 
Here,  ${s}$ is a scale parameter representing the width of each distribution. Its value will be made explicit once the distance is defined.

Each network has a Laplacian matrix and associated spectral function. 
It follows that a measure of the distance between two spectral functions can be interpreted as the distance between the networks themselves.  
Labelling the networks $1$ and $2$, the IM-distance is then defined by
\begin{equation}
\epsilon 
= \sqrt{ \int_{0}^{\infty} [ \rho_1(\omega) - \rho_2(\omega) ]^2 d \omega}.
\end{equation}
This vanishes when the two networks are identical [$\rho_1(\omega) = \rho_2(\omega)$]. 
The parameter $s$ acts as a scale and is set by considering a complete graph (of the same size as Network~1, say) and an empty graph (of the same size as Network~2). 
Choosing $s$ so that $\epsilon =1$ for these networks sets  the IM-distance to be $1$ when two networks are maximally different.
Keeping this particular value of $s$, we now evaluate the distance for the two networks of interest.

As the  networks considered here are all relatively sparse, low values of $\epsilon$ are expected. 
Nonetheless, they serve to indicate which networks are most similar and dissimilar to each other.
The results are recorded in Table~\ref{table5} as the IM-distances from \emph{Ossian} to each of the other texts. 
They indicate that the Ossianic networks are two or three times further from the Homeric counterparts  than they are to those in \emph{Acallam na Sen{\'{o}}rach} or Lady Gregory's text. 
The results for the giant components are given in 
Appendix~\ref{AppendixB}.

Therefore all three approaches -- based on parametric fits, Kolmogorov-Smirnov tests and IM distances -- deliver the same result; 
the social-network structure of {\emph{Ossian}} bears a measurably closer resemblance to those of the Irish corpus than to the Homeric narratives which Macpherson and his allies strove to parallel.

~\\
\section{Conclusions}
\label{Conclusions}

James Macpherson's {\emph{Ossian}} has had a profound and far-reaching influence on Western culture, acting as a catalyst for Romanticism in art and literature and drawing attention to national folklore, mythology, and poetry throughout Europe and, indeed, the world. 
However, it has also been referred to as a ``curious blend of misreading and willful misrepresentation''~\cite{Thomson1952}
and  ``the most successful literary fraud in history''~\cite{Curley2009}.
Ossianic scholarship has enjoyed a revival in recent decades, in part due to revisionist interpretations surrounding the production of the texts and  their contexts in the history of cultural colonization. 
As a result Macpherson has enjoyed a degree of rehabilitation and Ossianic scholarship has seen a growing emphasis on research questions that seek greater nuance than that found in the divisive polarities of truth and fraud~\cite{Gaskill2008}.

Macpherson employed paratexts -- footnotes, prefaces, and introductory dissertations -- to assert the value of the {\emph{Ossian}} poems in revealing a  supposed early history of Scotland and Ireland, and to argue for their consideration alongside recognised examples of ancient epic poetry. 
With the aid of Hugh Blair, he very deliberately positioned {\emph{Ossian}} within an established classical context in order to add legitimacy and authority to the new Scottish epic. 
Both used the same paratexts to repeatedly compare the poems to Greek and Latin prototypes -- to Homer in particular -- and to distance them from Irish sources, which were perceived to have slighter classical and cultural capital. 
Here, we have shown that network analysis exposes, to a certain extent, less immediately observable correlations between these sources -- {\emph{Ossian}}'s unelective affinities with other texts in the epic tradition.

By determining the statistics which describe the structural connectedness of the social networks present in {\emph{Ossian}} and those of {\emph{Ossian}}'s Irish and Greek epic intertexts, we are able to make meaningful quantitative comparisons. 
Whether we use the full or positive networks (or their giant components -- see Appendix~\ref{AppendixB}), 
there is a clear match between \emph{Ossian} and {\emph{Acallam na Sen{\'{o}}rach}} 
(as well as Lady Gregory's tales of the Fianna).
On the other hand,  there is little or no detectable structural resemblance  between  
{\emph{Ossian} and the works of Homer, whose graphs are also unlike those of the Irish texts.
Almost a quarter of a millenium ago, Charles O'Connor and Sylvester O'Halloran identified  characters  in the Ossianic network as appropriated from Irish tradition.
These characters are nodes in the social-network interpretation.
Here we have shown that the configuration of edges of the network -- the set of relationship links in the narrative -- also resembles that of Irish mythology, and not that of Homer.
In this sense our work complements that of O'Connor and O'Halloran.

These correlations, of course, do not imply causation. 
Thomson's extensive work~\cite{Thomson1952} in identifying {Irish} parallels for the narratives in the \emph{Ossian} corpus deepens our understanding of Macpherson's relationship to oral and written Gaelic culture {in both Ireland and Scotland}. 
However, it cannot resolve the binary debate about the authenticity or inauthenticity of \emph{Ossian}, or decide nebulous concerns about Macpherson's designs or intentions in shaping the work. 
Similarly, the structural resemblance of the social networks present in {\emph{Ossian} to those of the Irish epics do not necessarily prove that these were Macpherson's sources. 
Given that allusions to Homer are found in the paratexts, which are not included in the network analysis data, the outcome appears reasonable. 
What this analysis does reveal is a statistical picture at odds with the argumentative frame of \emph{Ossian}: a structural affinity with the Irish texts whose influence Macpherson explicitly disavows, and a dissonance from the Greek sources whose authority and eminence \emph{Ossian} seeks to emulate.

\vspace{1cm}
\noindent
{\bf{Acknowledgements:}} We would like to thank Silvano Romano for encouraging us to carry out this investigation.  
Thanks to Robin de Regt for help with gathering data in the early part of this project.
This work was supported by the EU FP7 Projects 
No. 269139, ``Dynamics and Cooperative Phenomena in Complex Physical and Biological Media,'' 
No. 295302, ``Statistical Physics in Diverse Realizations'' and
No. 612707, ``Dynamics of and in Complex Systems,'' 
and by the Doctoral College for the Statistical Physics of Complex Systems, Leipzig-Lorraine-Lviv-Coventry, Franco-German University. 
PMC is supported by a European Research Council (No. 295663) Advanced Investigator grant to R.I.M. Dunbar.


%

\newpage

\appendix
\section{Appendix: Individual Characters in {\emph{Ossian}}}
\label{AppendixA}

To construct the Ossianic network we use Gaskill's modern scholarly edition, {\emph{The Poems of Ossian and Related Works}}. This has five main sections: 
{\emph{Fragments of Ancient Poetry}} (1760); 
{\emph{Preface to}} Fingal (1761/62); 
{\emph{The Works of Ossian}} (1765): Fingal; 
{\emph{The Works of Ossian}} (1765): Temora; 
and 
{\emph{Preface to}} the Poems of Ossian (1773). 
 We use the material from 1765 only ({\emph{Fingal}} and {\emph{Temora}}) 
because 
these comprise the complete Ossianic corpus; 
the poems in {\emph{Fragments}} are reproduced and expanded in  {\emph{Fingal}}, 
with occasional modifications to the spelling of characters' names. 


\begin{table}[!b]
\caption{The most important characters of the Ossianic texts ranked according to their betweenness centrality, closeness, eigenvector centrality,  and degree.
We employ Gaskill's superscript notation to identify characters that share their names with others in the text.
}
\begin{center}
\resizebox{\textwidth}{!}{%
\begin{tabular}{|l|l|l|l|l|l|l|l|} \cline{1-5} \cline{7-8} 
             & Rank   & Betweenness       & Closeness         & Eigenvector        &   & Rank & Degree $( k )$   \\
						\cline{1-5} \cline{7-8} 
             &  1     & Fingal (0.38)     & Fingal (0.43)     & Fingal (0.53)      &   & 1    & Fingal (74)      \\ \multirow{4}{*}{\rotatebox{90}{Full}}    
             &  2     & Ossian (0.26)     & Ossian ({0.40})      & Ossian (0.48)      &   & 2    & Ossian (61)      \\
             &  3     & Cuchullin (0.12)  & Swaran (0.36)     & Gaul$^1$ (0.35)    &   & 3    & Cuchullin (30)   \\
             &  4     & Swaran (0.08)     & Carril (0.35)     & Fillan (0.32)      &   & 3    & Gaul$^1$ (30)    \\
             &  5     & Gaul$^1$ (0.06)   & Gaul$^1$ (0.35)   & Fergus$^1$ (0.17)  &   & 5    & Cairbar$^2$ (21)  \\
\cline{1-5} \cline{7-8}
            &  1      & Fingal (0.38)     & Fingal (0.40)     & Fingal (0.44)      &   & 1     & Fingal (68)      \\ \multirow{4}{*}{\rotatebox{90}{Positive}}
            &  2      & Ossian (0.26)     & Ossian (0.37)     & Ossian (0.38)      &   & 2     & Ossian (51)      \\
            &  3      & Cuchullin (0.12)  & Carril (0.33)     & Gaul$^1$ (0.21)    &   & 3     & Cuchullin (27)   \\
            &  4      & Swaran (0.08)     & Swaran (0.32)     & Fillan (0.19)      &   & 4     & Gaul$^1$ (20)    \\
            &  5      & Gaul$^1$ (0.06)   & Cuchullin (0.32)  & Fergus$^1$ (0.18)  &   & 5     & Connal$^5$ (18)  \\
\cline{1-5} \cline{7-8}
            &  1      & Cairbar$^2$ (0.34)& Cairbar$^2$ ({0.20}) & Gaul$^1$ (0.49)    &   & 1     & Cairbar$^2$ (10)\\ \multirow{4}{*}{\rotatebox{90}{Negative}}
            &  2      & Swaran (0.27)     & Ardan$^2$ ({0.20})   & Ossian (0.48)      &   & 1     & Gaul$^1$ (10)    \\
            &  3      & Ardan$^2$ (0.24)  & Swaran (0.19)     & Cremor$^1$ (0.24)  &   & 1     & Ossian (10)      \\
            &  4      & Gaul$^1$ (0.24)   & Fingal (0.18)     & Cathmin$^1$ (0.24) &   & 4     & Swaran (7)       \\
            &  5      & Fingal (0.17)     & Gaul$^1$ (0.18)   & Leth$^1$ (0.24)    &   & 5     & Fingal (6)       \\
\cline{1-5} \cline{7-8}
\cline{1-5} \cline{7-8} 
\end{tabular}
}
\end{center}
\label{tableA1}
\end{table}

The network statistics listed in Table~\ref{table1} are averages.  
Literary scholars may also find the statistics corresponding to individual characters in the Ossianic narratives of interest. 
To this end,  in Table~\ref{tableA1} we rank top characters according to various criteria.
The {\emph{betweenness centrality}} of a node is the number of geodesics that pass through that node \cite{Newman_book}. 
Thus, a node with high betweenness controls how information is transferred  through the network.
The sum of the distances of a node from all other nodes of a connected graph is termed its {\emph{farness}}. 
The reciprocal of farness  quantifies how central a node is and is termed its {\emph{closeness}} \cite{Newman_book}.
{\emph{Eigenvector centrality}} characterises node importance as a function of centralities of its neighbours and nodes are deemed influential according to how they are linked to other important nodes
\cite{Newman_book}.
A variant of eigenvector centrality is ``pagerank'' score used to rank websites.
Finally, the degree $k$ is explained in the main text.

In the index of names of Gaskill's text, superscripts are employed to identify characters that share a common name with other characters in the narative.
We employ Gaskill's notation in our identification of characters in Table~\ref{tableA1}.

\section{Appendix: The Giant Components}
\label{AppendixB}

\begin{figure*}[t]
\begin{center}
\includegraphics[width=0.49\columnwidth, angle=0]{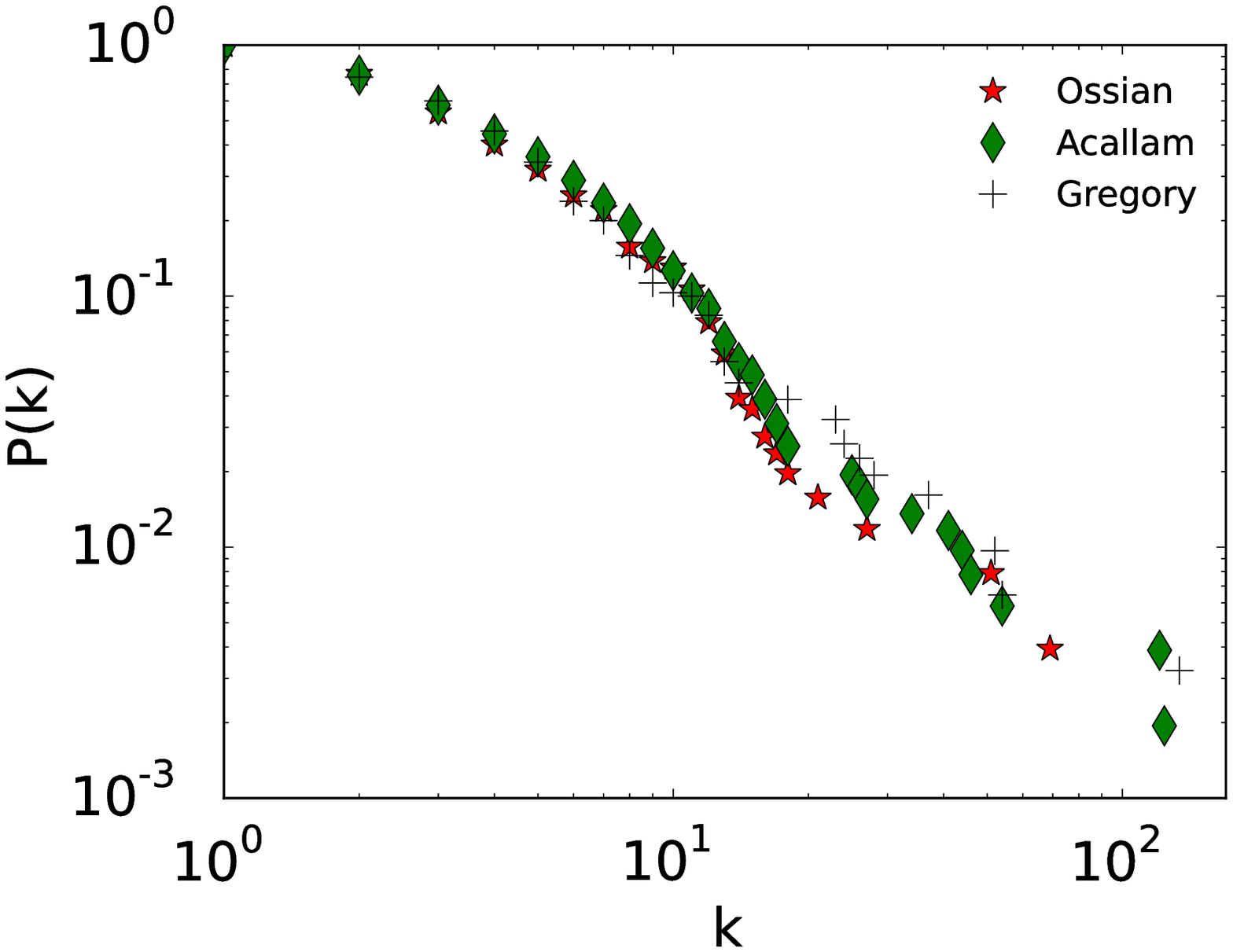}
\includegraphics[width=0.49\columnwidth, angle=0]{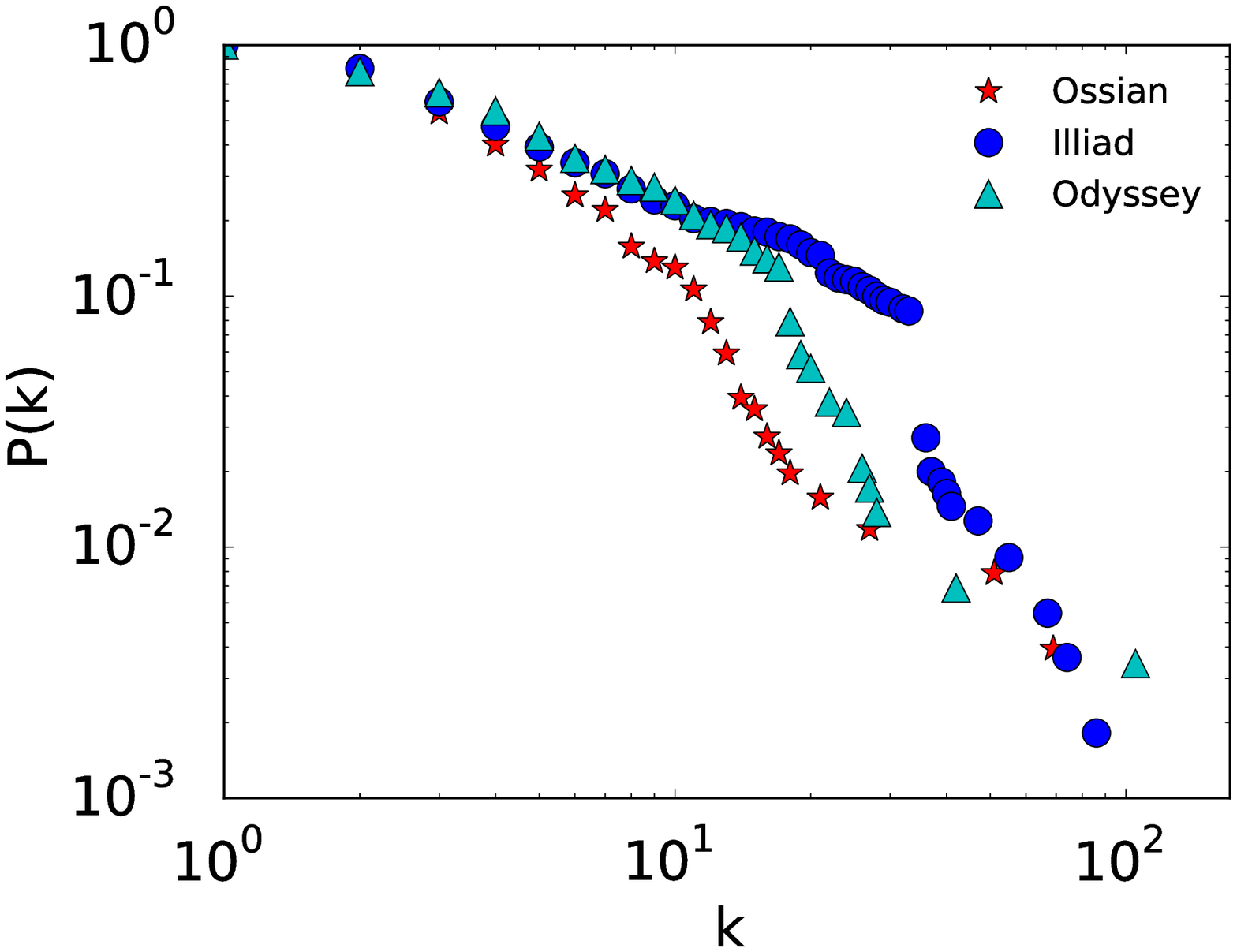}
\caption{As with Figure~2, the degree distributions for the giant components of the positive sub-networks depicted in {\emph{Ossian}} more closely resembles those of {\emph{Acallam na Sen{\'{o}}rach}} and  Lady Gregory's text (denoted here by ``{\emph{Gregory}}'') than the {\emph{Iliad}} or {\emph{Odyssey}}.
}
\label{FigBp1}
\end{center}
\vspace{0.3cm}
\end{figure*}

\begin{table}[!b]
\caption{Properties of the  giant components of the positive networks for {\emph{Ossian}} and comparative texts. }
\begin{center}
\begin{tabular}{|l|r|r|r|r|r|r|r|r|r|r|} \hline 
Narrative          & $N$ & $M$   & ${\langle{k}\rangle}$  & $ \ell$    & {$C$}& {$C_T$}  & $r_k$  \\ \hline
{\emph{Ossian}}    & 254 & 596   & 4.69                   & 3.62       & {0.45}& 0.30  & -0.10   \\
{\emph{Acallam}}   & 515 & 1364  & 5.30                   & 3.71       & {0.41}&0.20  & -0.13  \\
{\emph{Gregory}}   & 310 & 808   & 5.21                   & 3.24       & {0.46}&0.18  & -0.18   \\
{\emph{Iliad}}     & 550 & 2275  & 8.27                   & 3.78       & {0.48}&0.57  & 0.05    \\
{\emph{Odyssey}}   & 291 & 987   & 6.78                   & 3.38       & {0.46}&0.39  & -0.08   \\
\hline 
\end{tabular}
\end{center}
\label{tableB1}
\end{table}

In Table~\ref{table4} of the main text, we saw evidence for  matches between the social structures of  {\emph{Ossian}} and {\emph{Acallam na Sen{\'{o}}rach}}.
However, the test failed to match the full networks of \emph{Acallam na Sen{\'{o}}rach}} and Lady Gregory's text.
One may reasonably expect that these should match because  the former is acknowledged as a source for the latter and we should be able to detect this.
(The  approach in {Subsection~3.3} is more successful in this regard.)
One may  surmise that in constructing the {\emph{Complete Irish Mythology}}, Lady Gregory may have mostly been interested in characters taking part in the main action - in other words those who, in network terms, belong to the giant component. 
For these reasons we repeat part of the analyses of the main text for the giant components of the positive sub-networks.
We start with  Table~\ref{tableB1} where we list the network statistics.


In the main text we also established that the degree distribution of the social network underlying Macpherson's {\emph{Ossian}} is similar to those of  {\emph{Acallam na Sen{\'{o}}rach}} {{and}}   dissimilar to the {\emph{Iliad}} and {\emph{Odyssey}}.
This result is  interesting because Macpherson made concerted efforts to distance his work from the Irish corpus and, with the likely cooperation of Hugh Blair, attempted to establish comparability with Homer~\cite{Michell2012}.

\begin{table}[!b]
\caption{Relative minimum-information-loss probabilities for  the various degree distribution functions of the giant components of the positive networks. 
The probabilities for the most likely distribution for a given narrative is highlighted in boldface.}
\begin{center}
\begin{tabular}{|l|c|c|c|c|c|} \hline  
                     & Power law &Truncated          & Log normal      & Exponential & Weibull     \\
                     &           & power law         &                 &             &             \\
\hline
{\emph{Ossian}}      & $\sim 0$  &   0.17            &   {\bf{0.45}}   &   $\sim 0$  &  {0.38}        \\
{\emph{Acallam}}     & $\sim 0$  &   $\sim 0$        &   {\bf{0.98}}   &   $\sim 0$  &  0.02        \\
{\emph{Gregory}}     & $\sim 0$  &   $\sim 0$        &   {\bf{0.99}}   &   $\sim 0$  &  0.01         \\
{\emph{Iliad}}       & $\sim 0$  &   $\sim$ {\bf{1}} &   $\sim 0$      &   $\sim 0$  &  $\sim 0$  \\
 {\emph{Odyssey}}    & $\sim 0$  &   0.33            &   0.17          &   $\sim 0$  &   {\bf{ 0.50 }}     \\
 \hline  
\end{tabular}
\end{center}
\label{tableB2}
\end{table}

\begin{table}[!b]
\caption{Maximum likelihood estimates for the various parameters associated with the probability distributions fitted to the data coming from the giant components of the positive sub-networks.  }
\begin{center}
\resizebox{\textwidth}{!}{%
\begin{tabular}{ |l                         |c          |c         |c                   |c          |c                   |c             |c       |c|}      \hline
                                            & Power     &\multicolumn{2}{c|}{Truncated} & \multicolumn{2}{c|}{Log normal}&Exponential &\multicolumn{2}{c|}{Weibull }  \\
                                            & law       &\multicolumn{2}{c|}{power law} & \multicolumn{2}{c|}{          }&   & \multicolumn{2}{c|}{          } \\\hline
                                            & $\gamma$  & $\gamma$ &$\kappa$            & $\mu$     & $\sigma$           & $\kappa$      & $\beta$ & $\kappa$ \\
\hline
{\emph{Ossian}}  & {2.1(8)} & 1.2(1)  &  {12.4(8)}  &  {\bf{0.8(8)}}  &   {\bf{1.1(11)}}  &  {4.3(1)}		& {0.5(3)} &   {1.2(5)}  \\\hline
{\emph{Acallam}} & {1.9(5)} & 1.2(1)  & {15.0(6)}   &  {\bf{1.1(5)}}   &  {\bf{1.0(3)}}   &  {5.1(1)}   & {0.5(2)} & {1.5(3)}     \\\hline
{\emph{Gregory}}  & {1.9(6)} & {1.2(1)}  & {15.7(7)}   &  {\bf{1.2(6)}}   &  {\bf{1.0(3)}}   &   {5.1(1)}  & {0.5(3)} & {1.4(4) }    \\\hline
{\emph{Iliad}}   & {1.8(4)} & {\bf{1.3(0)}}  & {\bf{40.0(4)}}   &  {0.3(8)}  & {1.7(0)}  &   {8.5(1)  } & {0.4(1)} & {0.7(5)}     \\\hline
{\emph{Odyssey}} & {1.8(6)} & {0.6(1)} & {11.6(13) }  &  {1.5(7)}  &  {1.0(2)}  &   {6.9(2)}  & {\bf{0.7(5)}}  & {\bf{4.5(2)}}     \\
 \hline 
\end{tabular}
}
\end{center}
\label{tableB3}
\end{table}

In Figure~\ref{FigBp1} we present the equivalent plot for the giant components of the positive sub-networks. 
Table~\ref{tableB2} records the corresponding probabilities for  the various degree distribution functions.
The corresponding parameter estimates are given in  Table~\ref{tableB3}.
In  Table~\ref{tableB4} we display the results of the Kolmogorov-Smirnov tests.
The  match between {\emph{Ossian}} and {\emph{Acallam na Sen{\'{o}}rach}} that was identified in {Table~\ref{table4}} of the main text is evident here but this time there is also a match to {{Lady Gregory's text.}}
{{The}} expected match between the latter and {\emph{Acallam na Sen{\'{o}}rach}} is also evident.
There is still no evidence for similarity between {\emph{Ossian}}  and either of the Homeric texts. 
Comparing the results of Table~\ref{tableB4} with those of Table~\ref{table4}, then, it would appear that the giant components of the positive sub-networks deliver even stronger results than those of the full networks.

\begin{table}[t!]
\caption{Results, in terms of $p$ values, of the Kolmogorov-Smirnov tests for the giant components of the full and positive sub-networks. 
Large values of $p$ suggest similarities in degree distributions and are highlighted in boldface.
 }
  ~\\
  \centering
  \begin{tabular}{clllll}
\cline{1-2}
\multicolumn{1}{|c|}{{Giant}}          & \multicolumn{1}{l|}{{\emph{Ossian}}}   &                                     &                                      &                                    &  
 \\ \cline{2-3}
\multicolumn{1}{|c|}{{Components}}     & \multicolumn{1}{l|}{{\bf{0.30} [{\bf{0.72}}]}}  & \multicolumn{1}{l|}{{\emph{Acallam}}}&                                     &                                    & 
 \\ \cline{2-4}
\multicolumn{1}{|c|}{{of Full}}        & \multicolumn{1}{l|}{{\bf{0.07} $[{\bf{0.20}}]$}}& \multicolumn{1}{l|}{{\bf{0.75} [{\bf{0.64}}]}} & \multicolumn{1}{l|}{{\emph{Gregory}}}&                          & 
 \\ \cline{2-5}
\multicolumn{1}{|c|}{{[Positive]}}       & \multicolumn{1}{l|}{{$<10^{-2}$}}             & \multicolumn{1}{l|}{{$<10^{-5}$}}    & \multicolumn{1}{l|}{{$<10^{-2}$}}   & \multicolumn{1}{l|}{{\emph{Iliad}}}&  
 \\ \cline{2-6}
\multicolumn{1}{|c|}{{Networks}}& \multicolumn{1}{l|}{{$<10^{-2}$}}             & \multicolumn{1}{l|}{{$<10^{-2}$}}    & \multicolumn{1}{l|}{{$<10^{-3}$}}   & \multicolumn{1}{l|}{$\le10^{-2}$}  & \multicolumn{1}{l|}{{\emph{Odyssey}}} \\ 
\cline{2-6}
\hline   
\end{tabular}
\label{tableB4}
\end{table}

This conclusion is reinforced by comparing the spectral distances, which are listed for the giant components of the positive networks in Table~\ref{tableB5}, to those of the full networks listed in Table~\ref{table5}. 
In particular, the distances from the networks of  {\emph{Ossian}} to those of {\emph{Acallam na Sen{\'{o}}rach}}, which were {{0.03}} {{in Table~\ref{table5}  are reduced  to 0.01 and 0.02 here.} }

Thus we conclude again that there are strong resemblances between the social networks of {\emph{Ossian}} and those of the Fenian cycle of Irish mythology, confirming the relationship between the two. Moreover, Macpherson's networks are unlike those of Homer.

\begin{table}[h!]
\caption{ Spectral distances from the giant component of the positive sub-networks of {\emph{Ossian}} to counterparts in the other texts. Lower values indicate a greater degree of similarity.} 
\begin{center}
\begin{tabular}{|l|c|c|c|c|} \hline 
Distance from {\emph{Ossian}}  & {\emph{Acallam}} & {\emph{Gregory}}  & {\emph{Iliad}} & {\emph{Odyssey}} \\ \hline
 Giant Component & 0.01 & 0.02 & 0.07 & 0.08 \\ \hline
 Positive GC     & 0.02 & 0.03 & 0.09 & 0.08  \\ 
\hline  
\end{tabular}
\end{center}
\label{tableB5}
\end{table}


\begin{thebibliography}{99}


\bibitem{Macpherson1761}
Macpherson, J., 
\emph{Fingal, an Ancient Epic Poem, in six books; together with several other poems, composed by Ossian the son of Fingal; translated from the Galic language}
(T.~Becket and P.A.~De~Hondt, London, [December 1761] 1762).


\bibitem{Macpherson1763}
Macpherson, J., 
\emph{Temora, an Ancient Epic Poem, in eight books; together with several other Poems, composed by Ossian, the son of Fingal; translated from the Galic language}
(T.~Becket and P.A.~De~Hondt, London, [March] 1763).


\bibitem{Macpherson1765}
Macpherson, J., 
\emph{The Works of Ossian, the son of Fingal, in two volumes. Translated from the Galic language by James Macpherson. The third Edition. To which is subjoined a critical dissertation on the poems of Ossian. By Hugh Blair, D.D.}
(T.~Becket and P.A.~De~Hondt, London, 1765).


\bibitem{Macpherson1773}
Macpherson, J., 
\emph{The Poems of Ossian. Translated by James Macpherson, Esq., in two volumes.
A new edition, carefully corrected, and greatly improved}
(W.~Strahan and T.~Becket, London, 1773).


\bibitem{Gaskill1996}
Gaskill, H. (ed.), 
{\emph{The Poems of Ossian and Related Works}}
with an Introduction by Fiona Stafford
(Edinburgh University Press, Edinburgh, 1996).

\bibitem{Johnson1775} 
Johnson, S., 
{\emph{A Journey to the Western Islands of Scotland}}
(A.~Strahan \& T.~Cadell, London, 1775).

\bibitem{OHalloran1763}
O'Halloran, S., 
The poems of Ossine, the son of Fionne Mac Comhal, re-claimed,
{\emph{Dublin Magazine}} 1763 pp.~21-23; 
reprinted in 
Vol. III of Ref.\cite{Moore2004}
pp.~87-89.



\bibitem{Lyons2007} 
Lyons, C.E., 
Sylvester O'Halloran: Miso-Dolos,
{\emph{Journal of the Galway Archaelogical and Historical Society}}
{\bf{59}}  (2007) 46-58.


\bibitem{Warner1762}
Warner, F., 
\emph{Remarks on the History of Fingal, and other poems of Ossian: Translated by Mr Macpherson. a letter to the Right Honourable the Lord L}. 
(H.~Payne, W.~Cropley and J.~Walter, London, 1762).

\bibitem{Webb1763}
Webb, D., 
{\emph{Fingal Reclaimed}} [Pamplet] 
(The Author, London, 1762);
reprinted in 
Vol. III of Ref.\cite{Moore2004}, pp.~78-86.

\bibitem{OBrien1764} 
O'Brien, J. [under the pen name M. de C.],
{\emph{M{\'{e}}moire au suject des po{\"{e}}mes de M.~Macpherson}},
in Journal des S{\c{c}}avans {\bf{8}} (1764) 845-857.

\bibitem{MacKenzie1805}
Mackenzie, H. (ed.),  \emph{Report of the Committee of the Highland Society of Scotland appointed to inquire into the Nature and Authenticity of the Poems of Ossian}. 
(Edinburgh University Press, Edinburgh, 1805).

\bibitem{Mulholland2009}
Mulholland, J.,  
James Macpherson's Ossian poems, oral traditions, and the invention of voice,
{\emph{Oral Tradition}} {\bf{24}} (2009) 393-414.

\bibitem{Baer2012}
B{\"{a}}r, G. and Gaskill, H.,  
\emph{Ossian and National Epic}
(Peter Lang, Frankfurt am Main, 2012).

\bibitem{Burnett2011}
Burnett,~A. and Anderson,~L.,
{\emph{Blind Ossian's Fingal: Fragments and controversy a reprint of the first edition and abridgement of the follow-up with new material}}
(Luath Press Ltd., Edinburgh, 2011).

\bibitem{Curley2009}
Curley, T.M.,  
\emph{Samuel Johnson, the Ossian Fraud, and the Celtic Revival in Great Britain and Ireland}
(Cambridge University Press, Cambridge, 2009).

\bibitem{Meek1991}
Meek, D.E., 
The Gaelic ballads of Scotland: creativity and adaptation,
in Ref.\cite{Gaskill1991}, pp.~19-48.

\bibitem{Fielding1996}
Fielding, P., 
\emph{Writing and Orality: Nationality, culture, and nineteenth-century Scottish fiction} 
(Oxford University Press, Oxford, 1996).

\bibitem{Gaskill1991}
Gaskill, H. (ed.),
{\emph{Ossian Revisited}}
(University of Edinburgh Press, Edinburgh, 1991).

\bibitem{StaffordinGaskill1996}
Stafford, F.,
Introduction: The Ossianic poems of James Macpherson, 
in 
Ref.\cite{Gaskill1996}.

\bibitem{Gaskill2004}
Gaskill, H. (ed.),
{\emph{The Reception of Ossian in Europe}},
(Thoemmes Continuum, London, 2004).

\bibitem{Gaskill2008}
Gaskill, H.,
{\emph{The Homer of the North}},
Interfaces {\bf{27}} (2007-08)  13-24.

\bibitem{Kidd1993}
Kidd, C.,
\emph{Subverting Scotland's Past: Scottish Whig Historians and the Creation of an Anglo-British Identity 1689-1830}
(Cambridge University Press, Cambridge 1993).

\bibitem{McNeil2007}
McNeil, K., 
\emph{Scotland, Britain, Empire: Writing the Highlands, 1760-1860}
(The Ohio State University Press, Columbus, 2007).

\bibitem{Moore2000}
Moore, D., 
Heroic incoherence in James Macpherson's {\emph{The Poems of Ossian}},
{\emph{Eighteenth-Century Studies}} {\bf{34}} (2000) 43-59.


\bibitem{Moore2004}
Moore, D., 
{\emph{Ossian and Ossianism}} 
(Taylor and Francis, London, 2004). 

\bibitem{Stafford1991}
Stafford, F.J., 
`Dangerous Success': Ossian, Wordsworth, and English Romantic Literature
in 
Ref.\cite{Gaskill1991}~ pp.~49-72.

\bibitem{Womack1989}
Womack, P.,  
\emph{Improvement and Romance: Constructing the Myth of the Highlands},
(Macmillan, London, 1989).

\bibitem{awforum}
Mitchell, S. (ed.), {\emph{Ossian in the Twenty-First Century.}} 
Spec. issue of Journal for Eighteenth-Century Studies {\bf{39}}.2 (2016) 159-250.


\bibitem{ourEPL}
Mac~ Carron,~P. and Kenna,~R.,
Universal properties of mythological networks, 
\emph{EPL} {\bf{99}} (2012) 28002.

\bibitem{Macpherson1760}
Macpherson, J., 
{\emph{Fragments of ancient Poetry, Collected in the Highlands of Scotland, and Translated from the Galic or Erse Language}},
G.~Hamilton and J.~Balfour,  Edinburgh, 1760).

\bibitem{Quint1993}
Quint, D.,
{\emph{Epic and Empire: Politics and Generic Form from Virgil to Milton}},
(Princeton University Press, New Jersey, 1993).

\bibitem{Golden2002}
Golden, J.L. and Golden, A.L., 
\emph{Thomas Jefferson and the Rhetoric of Virtue},
Rowman and Littlefield (Maryland, 2002).

\bibitem{Johnson2016}
Johnson, A.L., 
Thomas Jefferson's Ossianic romance,
\emph{Studies in Eighteenth-Century Culture} 
{\bf{45}} (2016)  19-35.


\bibitem{Evans1764} 
Evans, E.,
{\emph{Some Specimens of the Poetry of the Ancient Welsh Bards. Translated into English,
with Explanatory Notes on the Historical Passages, and a Short Account of Men and Places mentioned by the Bards, in order to give the Curious some Idea of the Tastes and Sentiments of our Ancestors, and their Manner of Writing}},
(R. and J. Dodsley, London, 1764).


\bibitem{Skene}
Skene, W.F., 
\emph{The four Ancient books of Wales containing the Cymric Poems attributed to the Bards of the Sixth Century},
(Edmonston and Douglas, Edinburgh, 1868).

\bibitem{Percy1775}
Percey, T.,
\emph{Reliques of Ancient English Poetry: Consisting of old heroic ballads, songs, and other pieces of our earlier poets, (Chiefly of the lyric kind.) Together with some few of later date},
(J.~Dodsley, London, 1775).

\bibitem{Brooke1816}
Brooke, C., 
\emph{Reliques of Irish Poetry: Consisting of heroic poems, odes, elegies, and songs, translated Into English verse. With notes explanatory and historical, and the originals in the Irish character, to which is subjoined an Irish tale}. 
(J.~Christie, Dublin, 1816).


\bibitem{Michell2012}
Mitchell, S., 
Macpherson, Ossian, and Homer's Iliad, 
in
{\emph{Ossian and National Epic}},
ed. B{\"{a}}r,~G and  Gaskill~H.
(Peter Lang, Frankfurt am Main, 2012).


\bibitem{OConnor1766}
O'Connor, C.,
{\emph{Dissertations on the History of Ireland. To which is subjoined, a dissertation on the Irish colonies established in Britain. With some remarks on Mr. Mac~Pherson's translation of Fingal and Temora}} 
(G.~Faulkner, Dublin, 1766).

\bibitem{Halloran1989}
O'Halloran, C., 
Irish re-creations of the Gaelic past: The challenge of Macpherson's Ossian,
\emph{Past and Present} {\bf{124}} (1989) 69-95.

\bibitem{DooleyRow1999}
Dooley,~A. and Roe,~H.,
\emph{Tales of the Elders of Ireland: a new translation of Acallam na Sen{\' {o}}rach} 
(Oxford University Press, Oxford, 1999).

\bibitem{Iliad}
Homer, 
{\emph{The Iliad}}, 
originally translated by Rieu~E.V.,
revised and updated by Jones,~P. with Rieu,~D.C.H.
(Penguin Classics, London, 2003).

\bibitem{Odyssey}
Homer, 
\emph{The Odyssey},
translated by Shewring,~W., with an Introduction by Kirk,~G.S.
(Oxford University Press, Oxford, 1980).

\bibitem{LadyGregory}
Gregory, A.,
{\emph{Gods and Fighting Men: The Story of the Tuatha De Danaan and of the Fianna of Ireland, arranged and put into English by Lady Augusta Gregory}} 
(J. Murray, London, 1904);
reproduced in 
{\emph{Lady Gregory's Complete Irish Mythology}}
(Bounty Books, Vacaville, CA, 2004).

\bibitem{Newman2003}
Newman, M.E., 
The structure and function of complex networks, 
\emph{SIAM Review}, {\bf{45}} (2003) 167-256.

\bibitem{ourEPJB}
Mac~Carron, P. and Kenna, R., 
Network analysis of the {\'{I}}slendinga s{\"{o}}gur - the Sagas of Icelanders, 
{\emph{European Physical Journal B}} {\bf{86}} (2013) 407.



\bibitem{Fronczak}
Fronczak,A.,  Fronczak,~P. and Holyst~J.A., 
{\emph{Average path length in random networks}}
Physical Review E {\bf{70}} (2004) 056110. 


\bibitem{NewmanPark}
Newman, M.E. and Park, J., 
Why social networks are different from other types of networks,
\emph{Physical Review E} 
{\bf{68}} (2003) 036122.	


\bibitem{albert2002statistical}
Albert,~R. and Barab{\'a}si,~A.L., 
Statistical mechanics of complex networks, 
\emph{Reviews of Modern Physics}
{\bf{74}} (2002) 47-97.

\bibitem{PMCthesis}
Mac~Carron, P., 
{\emph{A Network Theoretic Approach to Comparative Mythology}},
PhD thesis, Coventry University, UK (2014). 


\bibitem{WattsStrogatz}
Watts, D.J. and Strogatz, S.H., 
Collective dynamics of `small-world' networks,
\emph{Nature} {\bf{393}} (1998) 440-442.
	

\bibitem{Newman2002}
Newman, M.E., Assortative mixing in networks,
\emph{Physical Review Letters} 
{\bf{89}} (2002) 208701.



\bibitem{Heider}
Heider,~F.,
{\emph{Attitudes and cognitive organization.}}
J. Psychol. {\bf{21}} (1946) 107-112.



\bibitem{Newman_power}
Newman, M.E., 
Power laws, Pareto distributions and Zipf's law,
\emph{Contemporary physics}
{\bf{46}} (2005) 323-351.


\bibitem{Amaral}
Amaral,~L.A.N., Scala,~A., Barth{\'{e}}l{\'{e}}my,~M. and Stanley,~H.E., 
Classes of small-world networks,
\emph{Proceedings of the National Academy of Sciences} {\bf{97}} (2000) 11149-11152.


\bibitem{Clauset2009}
Clauset,~A., Shalizi,~C.R. and Newman,~M.E.J., 
Power-law distributions in empirical data
\emph{SIAM Review}, {\bf{51}} (2009) 661-703.	



\bibitem{AIC}
Akaike, H., 
A new look at the statistical model identification,
\emph{IEEE Transactions on Automatic Control} {\bf{19}} (1974) 716-723.


\bibitem{Burnham}
Burnham,~K.P. and Anderson,~D.R.,
\emph{Model Selection and Multimodel Inference: A Practical Information-Theoretic Approach}
(Springer, New York, 2002).


\bibitem{conover1972kolmogorov}
Conover, W.J., 
A Kolmogorov goodness-of-fit test for discontinuous distributions,
\emph{Journal of the American Statistical Association} {\bf{67}} (1972) 591-596.


\bibitem{conover1980practical}
Conover, W.J.,  
\emph{Practical Nonparametric Statistics}
(Wiley, New York, 1980).


\bibitem{daniel1990applied}
Daniel, W.W., 
\emph{Applied Nonparametric Statistics}
(PWS-Kent, Boston, 1990).


\bibitem{Ipsen}
Ipsen, M. and Mikhailov, A.S.,  
\emph{Evolutionary reconstruction of networks},
Physical Review E {\bf{66}} (2002) 046109; 
Erratum, Physical Review E {\bf{67}} (2003) 039901.


\bibitem{Jurman}
Jurman,~G., Riccadonna,~S., Visintainer,~R. and Furlanello,~C.,  
Biological network comparison via Ipsen-Mikhailov distance,
arXiv:1109.0220.



\bibitem{Newman_book}
Newman, M., 
\emph{Networks: An Introduction}
(Oxford University Press, Oxford, 2010). 



\bibitem{Thomson1952}
Thomson,~D.S.,
{\emph{The Gaelic Sources of Macpherson's Ossian}}
(Oliver \& Boyd, Aberdeen, 1952).




\end{thebibliography}
\end{document}